\documentclass[journal,draftclsnofoot, onecolumn, 12pt,twoside]{IEEEtran}
\normalsize

\usepackage{graphicx}
\usepackage{fancyhdr}
\usepackage{graphicx}
\usepackage{times,amsmath,amsfonts}
\usepackage{amsmath,amssymb,amstext}
\usepackage{multirow}
\usepackage{hhline}
\usepackage{slashbox}
\usepackage{multirow}
\usepackage{cite}
\usepackage{epstopdf}
\usepackage{bm}
\usepackage{algorithm} 
\usepackage{algorithmicx}
\usepackage{algpseudocode}
\usepackage{color}
\usepackage[lofdepth,lotdepth]{subfig}
\usepackage{setspace}
\usepackage{array,booktabs,xcolor}
\usepackage[letterpaper, left=0.75in, right=0.75in, bottom=0.76in, top=0.76in]{geometry}
\newtheorem{lem}{\textbf{Lemma}}
\newtheorem{proposition}{\textbf{Proposition}}

\newtheorem{definition}{\textbf{Definition}}
\ifCLASSINFOpdf
\else
\fi

\begin{document}

\begin{spacing}{1.4}

% paper title
% can use linebreaks \\ within to get better formatting as desired
\title{Device-Clustering and Rate-Splitting Enabled Device-to-Device Cooperation Framework in Fog Radio Access Network }
%
%
% author names and IEEE memberships
% note positions of commas and nonbreaking spaces ( ~ ) LaTeX will not break
% a structure at a ~ so this keeps an author's name from being broken across
% two lines.
% use \thanks{} to gain access to the first footnote area
% a separate \thanks must be used for each paragraph as LaTeX2e's \thanks
% was not built to handle multiple paragraphs
%

\author{Md. Zoheb~Hassan,
        Md. Jahangir~Hossain,~\IEEEmembership{Senior Member,~IEEE,}
        %Mingbo~Niu,~\IEEEmembership{Student~Member,~IEEE,}
        %Cyril Leung,~\IEEEmembership{Member,~IEEE}
        Julian~Cheng,~\IEEEmembership{Senior Member,~IEEE,}
     and~Victor~C.~M.~Leung,~\IEEEmembership{Life Fellow,~IEEE} % <-this % stops a space
%\thanks{Manuscript received ; revised .}
\thanks{Md. Zoheb Hassan, Md. Jahangir Hossain and Julian Cheng are with School of Engineering, The University of British Columbia, Kelowna, BC, Canada (e-mail: zohassan@mail.ubc.ca, \{jahangir.hossain, julian.cheng\}@ubc.ca). Victor C. M. Leung is with the College of Computer Science, Shenzhen University, Shenzhen 518060, China, and also with the Department of Electrical
	and Computer Engineering, The University of British Columbia, Vancouver, BC, Canada (e-mail: vleung@ece.ubc.ca).}% <-this % stops a space
% <-this % stops a space
%\thanks{C. Leung is with the Department of Electrical and Computer Engineering, University of British Columbia, Vancouver, BC V6T 1Z4, Canada (e-mail:
%cleung@ece.ubc.ca).}% <-this % stops a space
\vspace{-0.4cm}
}
% note the % following the last \IEEEmembership and also \thanks - 
% these prevent an unwanted space from occurring between the last author name
% and the end of the author line. i.e., if you had this:
% 
% \author{....lastname \thanks{...} \thanks{...} }
%                     ^------------^------------^----Do not want these spaces!
%
% a space would be appended to the last name and could cause every name on that
% line to be shifted left slightly. This is one of those "LaTeX things". For
% instance, "\textbf{A} \textbf{B}" will typeset as "A B" not "AB". To get
% "AB" then you have to do: "\textbf{A}\textbf{B}"
% \thanks is no different in this regard, so shield the last } of each \thanks
% that ends a line with a % and do not let a space in before the next \thanks.
% Spaces after \IEEEmembership other than the last one are OK (and needed) as
% you are supposed to have spaces between the names. For what it is worth,
% this is a minor point as most people would not even notice if the said evil
% space somehow managed to creep in.

% The paper headers
\markboth{Submitted Paper}%
{Submitted Paper}
% The only time the second header will appear is for the odd numbered pages
% after the title page when using the twoside option.
% 
% *** Note that you probably will NOT want to include the author's ***
% *** name in the headers of peer review papers.                   ***
% You can use \ifCLASSOPTIONpeerreview for conditional compilation here if
% you desire.

% If you want to put a publisher's ID mark on the page you can do it like
% this:
%\IEEEpubid{0000--0000/00\$00.00~\copyright~2007 IEEE}
% Remember, if you use this you must call \IEEEpubidadjcol in the second
% column for its text to clear the IEEEpubid mark.

% use for special paper notices
%\IEEEspecialpapernotice{(Invited Paper)}

% make the title area
\maketitle

\setcounter{page}{1}
\vspace{-1.5cm}
\begin{abstract}
	\vspace{-0.5cm}
\textcolor{black}{Resource allocation is investigated to enhance the  performance of  device-to-device (D2D) cooperation in a fog radio access network (F-RAN) architecture.   Our envisioned  framework enables two  D2D links to share certain orthogonal radio resource blocks (RRBs) by forming device-clusters.  In each device-cluster, both  content-holder device-users (DUs) transmit to the content-requester DUs via an enhanced remote radio head  (eRRH)  over  the same RRBs. Such RRBs are shared with the uplink F-RAN as well.  The intra device-cluster interference is mitigated by exploiting both uplink and downlink rate-splitting schemes, and the inter  device-cluster interference is mitigated by using an orthogonal RRB allocation strategy.   Our objective is to maximize the end-to-end sum-rate of the device-clusters while reducing the interference between  D2D cooperation and the uplink F-RAN over the shared RRBs. Towards this objective,  a joint optimization  of device-clustering, transmit power allocations, assignment of device-clusters to the eRRHs, and allocation of RRBs among the device-clusters is presented. Since the joint optimization is NP-hard and intractable, it is decomposed into device-clustering and resource allocation  sub-problems, and efficient solutions to both sub-problems are developed. Based on the solutions to the sub-problems, a semi-distributed and convergent algorithm, entitled \textbf{r}ate-\textbf{s}plitting for \textbf{m}ulti-hop \textbf{D}2D (RSMD),  is proposed to obtain the device-clusters and resource allocation for these device-clusters. Through extensive simulations, efficiency of the proposed RSMD algorithm  over several benchmark schemes is demonstrated.}
\end{abstract}
\vspace{-0.65cm}
\section{Introduction}
Fog radio access network (F-RAN) improves the performance of the conventional cloud RAN (C-RAN)  architecture by introducing signal processing and computation capability at the remote radio heads (RRHs), and thus, F-RAN is a key enabling architecture of the beyond 5G   wireless networks \cite{F_RAN_Peng,F_RAN_Network_Rev_1,F_RAN_NOMA_Recent}.  The latency and burden over fronthaul in an F-RAN architecture can be remarkably improved by introducing cache-enabled device-to-device (D2D) networking, and the combined system is referred as the \textit{D2D-integrated F-RAN} \cite{D2D_FRAN_Access_1,Saif_TWC}.  \textcolor{black}{The achievable data rate of a D2D-integrated F-RAN is reduced severely  by the interference among the cache-enabled D2D links, and consequently, an interference management is crucially important to enable massive D2D cooperation in an F-RAN. However, the problem of interference management for D2D-integrated F-RAN  is not thoroughly investigated in the literature. To fill this void, this work aims to investigate the interference management problem for the D2D-integrated F-RAN architecture from  a resource allocation perspective. Specifically, clustering of the cache-enabled D2D links is a promising way to improve utilization of  the limited radio resource blocks (RRBs) in  an F-RAN architecture. Meanwhile,  by using the advanced network edge computation resources, decentralized interference management can significantly improve scalability of the resource allocation in an F-RAN with  the reduced latency.  We are therefore motivated to propose a decentralized interference management scheme while taking  device-clustering into account.}

\vspace*{-0.4cm}
\subsection{Related Works}
\textit{Related works in D2D resource allocation:} \textcolor{black}{Cooperative D2D networking  with the static or mobile relay is an effective solution to  improve the coverage of the D2D links and the power consumption at the mobile devices. Various resource allocation problems were studied  for cooperative D2D networking \cite{D2D_Relay_5, Vincent_D2D,D2D_Relay_1, D2D_C-RAN}. However, the aforementioned studies did not focus on the interference management issue of the cooperative D2D networks. In particular, an interference-free solution can be developed by scheduling each D2D link in the system over the orthogonal RRBs \cite{D2D_Relay_5, Vincent_D2D}. Alternatively, in contrast to the interference-free solution, one can consider spectrum sharing for a large number of D2D links while treating interference as a noise \cite{D2D_C-RAN,D2D_Relay_1}.  We emphasize that from an information theory perspective, neither approaches provide an optimal interference management \cite{Anas}. In this work, considering cooperative D2D networking,  we envision a resource allocation framework  that strikes a balance between the aforementioned two approaches by allowing certain interference and canceling the remaining interference of the system with the goal of improving the overall utilization of the RRBs.}

\textcolor{black}{Recently, power-domain non-orthogonal multiple-access (PD-NOMA), which allows more than one user over a single RRB,  has been introduced to manage interference in the D2D communication enabled network.  The sum-rate of D2D communications can be improved by forming D2D groups of one device-transmitter and two device-receivers and applying PD-NOMA in each D2D group \cite{D2D_NOMA}. 
%Moreover, the spectrum efficiency of a PD-NOMA enabled cellular network can be improved by reusing the spectrum  at the D2D network and mitigating the co-channel interference \cite{D2D_NOMA_2}. 
PD-NOMA is  also promising for a  cooperative D2D network where both cellular and D2D links share the same spectrum and device-transmitter of the D2D link works as a relay for the cellular users.  
Specifically,  PD-NOMA enabled cooperative D2D networking improves both the coverage and security of the cell-edge user  \cite{D2D_NOMA_3, D2D_NOMA_4}. 
Moreover, one can consider PD-NOMA to enhance both outage probability and average sum-rate  of a  cooperative relay sharing (CRS) system, where two users or devices communicate with their corresponding destinations simultaneously over the same RRB by using a common relay \cite{D2D_NOMA_5,D2D_NOMA_6}.  In this work, we envision the D2D-enabled cached-content distribution in F-RAN where multiple  D2D links  share  both RRB(s) and relay(s) for simultaneous data transmission or reception. Essentially, the envisioned D2D cooperation framework is a generalized CRS  system.  Moreover,  we consider that the D2D links reuse the cellular RRBs, and as a result,  performance of the envisioned  CRS system also depends on the mitigation of interference between the cellular and D2D links.  However, the optimality of PD-NOMA for such a generalized CRS system is not evident from the literature.    Accordingly, it is imperative to develop an interference-aware resource allocation scheme to accommodate multiple relay-assisted D2D links  over the same RRB(s) for data transmission and reception.}

\textit{Related works in rate-splitting:} \textcolor{blue}{Rate-splitting (RS) based transmission scheme can accommodate multiple users over the same RRB by mitigating interference. The effectiveness of RS  is confirmed from the classical information theory. In particular, the optimal rate-region of the two-user interference channel can be achieved by splitting each user’s message into common and private parts, and decoding the intended private and both common messages jointly at each user \cite{RS_Info_1}.  Meanwhile, the combination of RS and successive decoding can also achieve the entire capacity region of a Gaussian multiple access (MAC)  channel \cite{RS_Uplink}.  In the seminal work \cite{RS_5G}, the one-layer RS was proposed as a promising solution to mitigate interference and improve the capacity of a multiple-input multiple-output wireless network.  In the one-layer RS enabled transmission scheme, each user's message is split into a common part and a private part. All the common parts are combined into one  common message. Note that the common message is constructed by using a codebook shared by all users, and hence the common message is  decodable by all users. In contrast, each private message is decodable only by a specific user. Each user, therefore, first decodes the common message and removes its interference by applying the successive interference cancellation (SIC) technique, and subsequently, decodes the individual private message. Essentially, the RS based transmission partially decodes the interference and partially treats interference as noise. In other words, such an RS based transmission strikes a suitable balance between two extreme approaches of fully decoding the interference and fully treating interference as noise.  The effectiveness of RS for mitigating interference in the multiple-input single-output (MISO) broadcast (BC) channel was extensively studied \cite{RS_new_1,RS_New_2,RS_CMD_Org,RS_NOUM,RS_CMD_Review,RS_CMD_MISO,RS_DPC}. Specifically,  in the presence of imperfect channel state information at transmitter (CSIT), the one-layer RS not only improves the sum-rate compared to the conventional approach of treating interference as noise but also achieves the optimal degrees-of-freedom region of a MISO BC channel \cite{RS_new_1,RS_New_2}.  Moreover, the aforementioned one-layer RS strategy was generalized to a  powerful multiple access framework, namely, rate-splitting multiple access (RSMA), to improve both the spectral and energy efficiency of downlink communication system \cite{RS_CMD_Org,RS_NOUM}.  Particularly, by leveraging the linear precoding at the transmitter and SIC at the receivers,  RSMA outperforms the contemporary multiple access schemes for the multi-antenna network \cite{RS_CMD_Review}.  For a two-user MISO BC channel, RSMA encompasses the advantages of the well-known multiple access schemes, namely, PD-NOMA, space division multiple access, orthogonal multiple access, and multicasting \cite{RS_CMD_MISO}.  In addition, RSMA  achieves larger rate-region than the dirty-paper-coding (DPC) scheme for the MISO BC channel having imperfect CSIT, and the achievable rate-region of the MISO BC channel with imperfect CSIT can be further enhanced by using DPC to encode the private messages of RSMA scheme \cite{RS_DPC}.}

\textcolor{blue}{RSMA is also efficient for mitigating interference in the single-input single-output (SISO) network. In particular, the optimal power and rate allocations were derived to maximize the sum-rate of RSMA over the downlink SISO BC and uplink SISO MAC channels \cite{RS_CMD_1,Walid_Sadd_Uplink_RSMA,RSMA_NOMA}. Both RSMA and PD-NOMA achieve the same sum-rate for the SISO BC channel with two users. However, RSMA can outperform PD-NOMA for the SISO BC channel with more than two users, especially for the large SIC detection thresholds and high minimum data rate demands of the users \cite{RS_CMD_1}. We emphasize that such a case happens due to the intrinsic requirements of PD-NOMA for performing multiple SIC operations at certain users to fully decode the interference signals. Moreover, because of requiring only $1$ SIC operation at each user, RSMA   reduces the receiver complexity considerably than PD-NOMA \cite{RS_CMD_Org}. Meanwhile, unlike PD-NOMA, RSMA  achieves the entire capacity region of  the  uplink SISO MAC channels and thus, always outperforms PD-NOMA in terms of the achievable uplink sum-rate \cite{Walid_Sadd_Uplink_RSMA,RSMA_NOMA}.}

\textcolor{black}{Rate-splitting with common message decoding (RS-CMD) is another novel RS technique that was originally proposed to effectively mitigate interference in a downlink C-RAN architecture \cite{RS_CMD_Anas, RS_CMD_2,RS_CMD_3}.  In RS-CMD, a user can receive multiple common  messages from multiple RRHs, and the user decodes these messages  by applying multi-layer SIC operations.  It is therefore important to determine the set of RRHs that transmit common and private messages for each user, and the group of users who decode each common message. To this end, several heuristic algorithms were proposed \cite{RS_CMD_Anas, RS_CMD_2,RS_CMD_3}. 
	However, the proposed algorithms for  clustering users to decode each common message in an RS-CMD system require global channel state information (CSI) at the cloud processor \cite{RS_CMD_Anas, RS_CMD_2}, and consequently, these algorithms are not scalable.  The scalability of RS-CMD algorithm   can be improved significantly by considering  only statistical CSI for clustering the users to decode each common message in a large-scale C-RAN. Moreover, the statistical CSI based RS-CMD algorithm provides robustness against imperfect CSI in a C-RAN, and  improves the capacity of the conventional single-layer RS schemes \cite{RS_CMD_3}.}

Inspired by the effectiveness of RS for managing interference in both C-RAN and multi-user cellular network \cite{RS_CMD_Org, RS_NOUM,RS_CMD_Review, RS_CMD_MISO,  RS_DPC, RS_CMD_1,Walid_Sadd_Uplink_RSMA,RSMA_NOMA,RS_CMD_Anas,RS_CMD_2,RS_CMD_3}, we are motivated to exploit the advantage of RS to schedule multiple D2D links over the same RRB in an F-RAN architecture.  \textcolor{blue}{Note that  the RRHs in C-RAN have only the basic forwarding and RF processing capability, and consequently, C-RAN entirely uses the centralized signal processing and radio resource management. In contrast, the RRHs in an F-RAN have certain cloud computing capability, and such RRHs are referred as the eRRHs \cite{Saif_TWC}. Unlike RRHs, the eRRHs can perform local signal processing and radio resource management tasks. Thus, unlike C-RAN, an F-RAN  exploits both the centralized and edge processing for radio resource management, and thereby,  reduces latency for the end-users. However, to facilitate the local signal processing and radio resource management with the  reduced complexity, it is usually assumed that a given user or device is allowed to be connected with maximum one eRRH in F-RAN \cite{F_RAN_NOMA_Recent,F_RAN_Megumi_Conf_1,Megumi_EE_FRAN_1,Saif_Com_letter_1}.  Due to the aforementioned connectivity constraint, the performance gain of RS in an F-RAN will depend on the association between devices and eRRHs as well.} To this end, we optimize the performance of RS strategy in an F-RAN while solving an intricate problem of  association among devices, eRRHs, and  RRBs. We emphasize that compared to the existing RS literature, our work applies RS to an entirely new system setting.

\textit{Related works in the device-clustering}: \textcolor{black}{To take the advantage  of RS in the D2D-integrated F-RAN architecture, it is imperative to cluster  the D2D links that will share the same RRBs for data transmission and reception.  In the literature, the user-clustering problem was investigated for the PD-NOMA  systems.  The users in a downlink PD-NOMA system can be clustered by applying the heuristic methods, namely, distance-based user ranking, coalition-game, and matching-theory \cite{Distance_based_Clustering, Collation,Matching_1}.  In a multi-antenna PD-NOMA system, the $K$-means clustering and expectation-maximization based unsupervised-learning techniques can be used  to form the user-clusters  \cite{ML_Clustering, ML_clustering_2}. However, the aforementioned studies have certain limitations. Specifically, the existing learning-based user-clustering methods are  centralized \cite{ML_Clustering, ML_clustering_2}, and they require the global CSI of all the device-users (DUs) at the central server. As a result, the training overhead to obtain  the required data set is significantly increased making the centralized learning-based user-clustering algorithms impractical for a large-scale network.  Moreover, because of requiring the exhaustive search, an optimal clustering algorithm is  infeasible for a practical system. Accordingly,  we  propose a low-complexity algorithm for clustering the D2D links while reducing the high  training overhead requirements of the contemporary centralized device-clustering methods.}

\vspace*{-0.4cm}
\subsection{Contributions}
 \textcolor{black}{The motivation of this work is to develop resource allocation to enhance the performance of D2D cooperation in an F-RAN  by managing interference in the system efficiently. We propose an optimization framework by taking the cooperative D2D networking, device-clustering, and RS-based transmission into account. To the best of the authors' knowledge, this is the first work that presents an in-depth resource optimization for managing interference  in a D2D-integrated F-RAN architecture.} The main contributions of this work are summarized as follows:

\begin{enumerate}
	\item We propose clustering of the D2D links to use the orthogonal RRBs efficiently for data transmission and reception.  In each device-cluster,  both content-holder (CH) DUs transmit to the content-requester (CR) DUs via an eRRH  over the same RRBs. Such RRBs are also shared with the uplink F-RAN. \textcolor{black}{Essentially, our proposed framework enhances  utilization of the RRBs at the cost of interference.  To combat the inter device-cluster interference,  the device-clusters are scheduled over orthogonal RRBs; to combat the intra device-cluster interference, uplink and downlink RS are employed at each device-cluster;  and to combat the interference between D2D network and uplink F-RAN, RRB pricing-aware transmit power allocation is employed. Our objective is to maximize the end-to-end sum-rate of the device-clusters while reducing the interference between  D2D cooperation and the uplink F-RAN over the shared RRBs. Towards this objective,  a joint optimization  of device-clustering, transmit power allocations, assignment of device-clusters to the eRRHs, and allocation of RRBs among the device-clusters is presented. Since the joint optimization is NP-hard and intractable, we obtain an efficient solution by solving two sub-problems.}
	
	\item \textcolor{black}{In the first sub-problem, we  partition the D2D links into non-overlapping device-clusters.  Inspired by unsupervised learning, we propose a device-clustering method by applying two-dimensional principle component analysis (2D-PCA).  Thereby, a novel device-clustering algorithm is proposed. Our proposed device-clustering algorithm alleviates the burden of global CSI acquisition at the CBS,  and thus, reduces the training overhead in the system. Moreover,  our proposed device-clustering algorithm reduces computational complexity significantly compared to an exhaustive-search based device-clustering method with small performance loss.}
	
	\item In the second sub-problem,  we obtain resource allocation among the  device-clusters by solving a Stackelberg game. Here, the device-clusters, acting as  the followers,  solve the RRB pricing-aware power 
	allocation problem by using a non-cooperative potential game. The CBS, acting as a leader, solves the problem of scheduling the eRRHs and RRBs  among the device-clusters  and determines prices of the RRBs to control uplink interference. Novel and efficient solutions to the followers' and leader's sub-games are obtained by applying the  fractional programming and alternating optimization techniques, respectively.
	
	\item A semi-distributed algorithm, entitled \textbf{r}ate-\textbf{s}plitting for \textbf{m}ulti-hop \textbf{D}2D (RSMD), is proposed to obtain the device-clusters and resource allocation among the device-clusters. It is proved that the proposed RSMD algorithm converges to the Stackelberg-equilibrium (SE) resource allocation with polynomial computational complexity.
	Extensive simulations are conducted to verify the effectiveness of  the proposed RSMD algorithm over several benchmark schemes.

\end{enumerate}

The rest of the paper is organized as follows. Section II provides the overall system model and problem formulation. The  device-clustering and Stackelberg resource allocation game are discussed in Sections III and IV, respectively. Section V presents  the proposed RSMD algorithm. The numerical results and the concluding remarks are presented in Sections VI and VII, respectively.
%Several illustrative simulation results and the concluding remarks are presented in Sections VI ad VII, respectively. 
%%Due to the space limitation,  proofs of the lemmas and propositions  are provided in the supplementary materials.

\textit{Notations:} $\mathcal{X}$, $\mathbf{X}$, and $\mathbb{X}$ represent a set, a vector, and a matrix, respectively;  $\mathbf{X}[i]$ denotes the $i$-th element of the vector $\mathbf{X}$; $|\mathcal{X}|$ denotes the cardinality of a set  $\mathcal{X}$; $\mathbf{X}^T$ and $\mathbb{X}^T$ denote the transpose of the vector $\mathbf{X}$  and the matrix $\mathbb{X}$, respectively; $\mathbb{E}[\cdot]$ is the expectation operator; $\left|\left|\cdot\right|\right|_2$ is the Euclidean norm; $\geqq$ is the element wise inequality; and $\mathbb{R}^{a \times b}$ is a real matrix of $a\times b$ dimension.

  \setlength{\textfloatsep}{0pt}
\begin{figure}
	%  	\vspace{-0.2cm}
	\begin{center}
		\label{fig_1}
		\includegraphics[width=0.7\linewidth, draft=false]{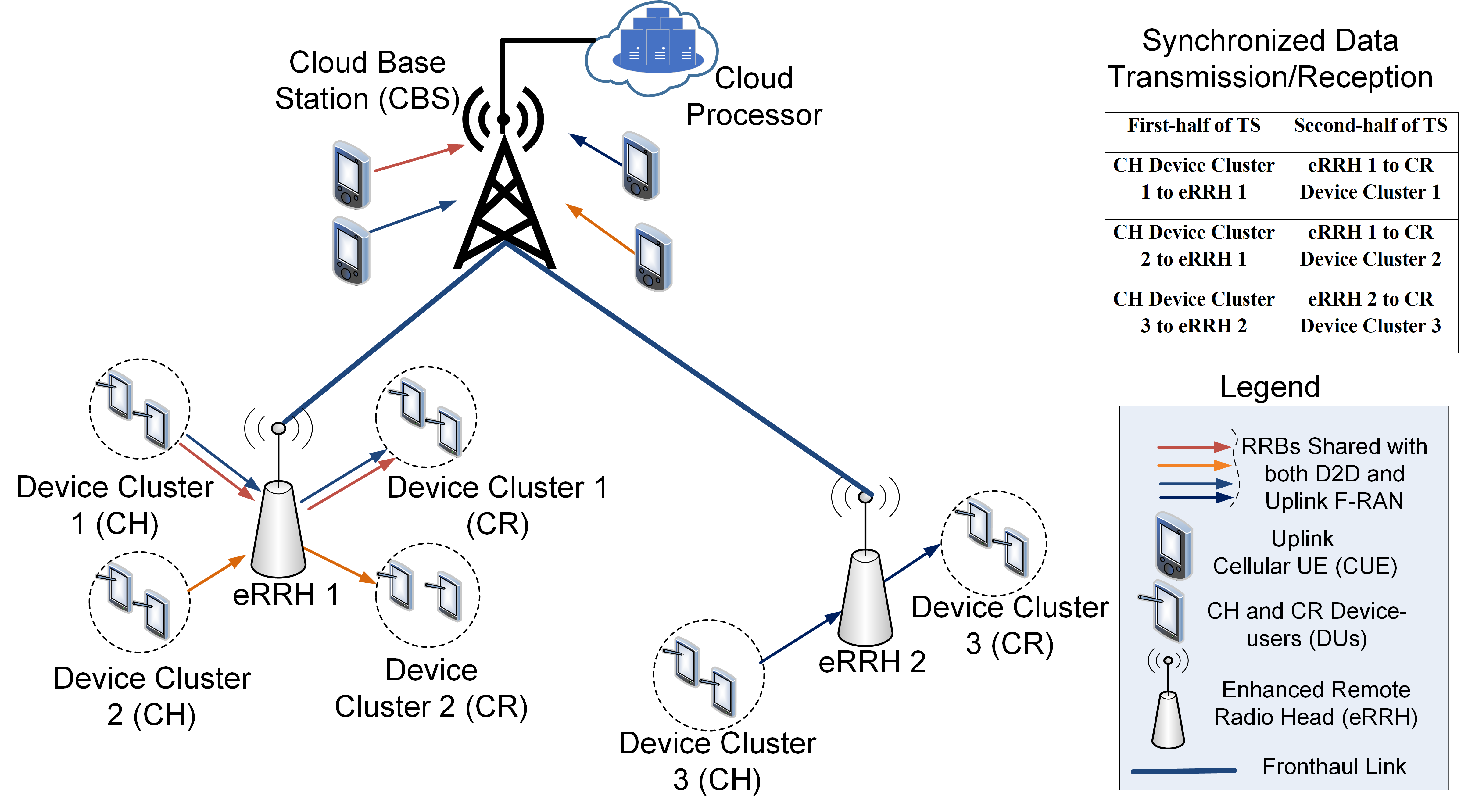}
		\caption{A cluster-based D2D cooperation  topology in F-RAN}
	\end{center}
	%  	\vspace{-1.4cm}
\end{figure}

\vspace{-0.2cm} 
\section{System Model and Problem Formulation}
\vspace{-0.2cm}
\subsection{System Overview}
\vspace{-0.1cm}
Consider an F-RAN with D2D cooperation, depicted in Fig. 1, with a CBS, $N$ non-overlapping RRBs, $C$ cellular UEs (CUEs) transmitting to the CBS in uplink, $M$ D2D links, and $L$ eRRHs. Each D2D link has one CH DU and  one CR DU where the CH DU and CR DU are the transmitting and the receiving devices, respectively.  
%CH DU transmits certain data file to the CR DU by using the available RRBs. 
We consider that in such  D2D links,  CR DUs are far from their corresponding CH DUs and  a relay  is required to assist the D2D communication. In this regard, the eRRHs are utilized to relay the data  from  CH DUs to CR DUs. The eRRHs can perform encoding, decoding,  and local radio resource allocation functionalities \cite{F_RAN_NOMA_Recent}.  Let, $\mathcal{L}=\{1,2, \cdots,L\}$ be the set of available eRRHs, $\mathcal{M}=\{1,2,\cdots,M\}$ be the set of D2D links, and  $\mathcal{N}_{sc}=\{1,2,\cdots, N\}$ be the set of available non-overlapping RRBs. Both DUs and eRRHs are half-duplex, and they have single antenna.

To utilize the given RRBs for the D2D links  efficiently, we adopt clustering of the  D2D links. To simplify both the encoding and decoding operations, we consider that each device-cluster contains two D2D links. The device-clusters are pairwise disjoint, and the total number of device-clusters in the network is $\lceil\frac{M}{2}\rceil$.  
Each device-cluster is allocated a set of non-overlapping orthogonal RRBs, and these RRBs are also shared with certain CUEs for uplink data transmission. Particularly, the assignment of the RRBs among the CUEs is predefined, and our proposed scheme controls the transmit power of both CH DUs and eRRHs as such  interference to the uplink cellular links remains small. Both CH DUs and CR DUs in a given device-cluster transmit and receive, respectively, over the allocated RRBs by using an RS strategy.   In this setup, both CH DUs of a given device-cluster  first transmit to an eRRH, and then, the eRRH forwards the received data to both CR DUs of that device-cluster. Each device-cluster is assisted by a single eRRH whereas a given eRRH  can assist multiple device-clusters. For the analytical tractability, the following assumptions are made. \textbf{A1:} The described system is operated on a slotted-time basis where the overall time duration is divided into equal and non-overlapping time-slots (TSs). Moreover, data transmission and reception in each TS are synchronized. As shown in Fig. 1, in the first-half of each TS,   the CH DUs of the device-clusters transmit data to the selected eRRHs, and then, the eRRHs forward the received data to the corresponding CR DUs in the second-half of the TS. 
%An example scenario considering three device-clusters and two relay F-UEs is depicted in Fig. 1.
The aforementioned synchronization can be achieved by sending timing-signals from the CBS to the DUs and eRRHs over the dedicated control channels  \cite{Synchronization}. 
\textbf{A2:} A slowly changing network topology is considered  where locations of the DUs  remain fixed for a number of TSs. \textbf{A3:} To facilitate power allocations, accurate local CSI is assumed to be available at each device-cluster \cite{Megumi_EE_FRAN_1}, \cite{Hayseem_NOMA_FRAN}. 

\textcolor{blue}{\textit{Remark 1:} The proposed scheme improves the performance of the cache-enabled D2D networking by leveraging both the computing and communication resources of an F-RAN.  The intrinsic property of an F-RAN is that the F-RAN exploits both the centralized cloud computing and local processing at the network edge \cite{F_RAN_Peng,F_RAN_Network_Rev_1,F_RAN_NOMA_Recent}.  We emphasize that  the proposed scheme has certain features that comply  with the notion of joint edge-cloud processing  capability of the standard F-RAN architecture.  First, the proposed scheme uses the computation capability of both the DUs and eRRHs, and distributes the radio resource allocation tasks between the eRRHs and CBS. Second, the proposed scheme allows each device-cluster to be served by only one eRRH, and thus, exploits the  local edge processing capability of an F-RAN with the reduced complexity. Third,  the proposed scheme utilizes the signal processing capability of the eRRHs for  decoding and encoding  of the  data transmitted by CH DUs.  Hence,  the fronthaul links do not need to carry the user data from the eRRHs to CBS, and consequently, the burden over the fronthaul links is alleviated significantly. Finally, the proposed scheme requires only the local CSI at the eRRHs and does not require the global CSI  at the CBS. Note that the aforementioned features are inherent to both the standard F-RAN network \cite{F_RAN_Megumi_Conf_1,Megumi_EE_FRAN_1,Saif_Com_letter_1,Hayseem_NOMA_FRAN}  and the D2D-integrated F-RAN system \cite{D2D_FRAN_Access_1,Saif_TWC}. Consequently, our proposed scheme is suitable for an F-RAN network.}

\vspace*{-0.35cm}
\subsection{Rate-splitting Enabled Transmission}
\vspace*{-0.1cm}
Without loss of generality, we describe the RS enabled transmission in the $j$-th device-cluster. The $j$-th device-cluster is denoted by $\mathcal{S}_j=\{\mathcal{S}_{j,1}, \mathcal{S}_{j,2}\}$ where $\mathcal{S}_{j,1}$ and $\mathcal{S}_{j,2}$  denote the first and second  D2D links belong to the $j$-th device-cluster, respectively. We assume that the CH DUs of the $j$-th device-cluster transmit to  the CR DUs via the $l$-th eRRH over the $n$-th RRB. \textcolor{black}{An overview of the considered RS strategy in the first and second hops of each device-cluster is given as follows. Note that the transmission from the CH DUs to eRRH is an uplink RS scheme. As per \cite[Lemma 3]{Walid_Sadd_Uplink_RSMA}, only one CH DU needs to split its message. In accordance with \cite{RSMA_NOMA}, the  CH DU having (relatively) superior channel gain splits its message into two streams and transmits a superposition-coded signal. On the other hand, the CH DU with  (relatively) weak channel gain, transmits a single stream. Without loss of generality, we assume that the CH DU  of the  $\mathcal{S}_{j,1}$ D2D link has a better channel gain with the $l$-th eRRH over the $n$-th RRB.  Hence, the CH DU of  the  $\mathcal{S}_{j,1}$ D2D link transmits two streams, namely, $s_{j,1}^{(1)}, s_{j,2}^{(1)}$, and  the CH DU of  the  $\mathcal{S}_{j,2}$ D2D link transmits a single stream $s_{j}^{(2)}$.  By applying SIC,  the $l$-th eRRH  decodes the received data streams in the order of $s_{j,1}^{(1)} \rightarrow s_{j}^{(2)}\rightarrow s_{j,2}^{(1)}$ \cite{RSMA_NOMA}.  In contrast, the transmission from  eRRH to the CR DUs is a downlink RS scheme. Hence, the $l$-th eRRH, splits the received message into a common and two private messages, and transmits a superposition-coded signal. The $l$-th eRRH assigns certain power and rate to the common message as such both  CR DUs can decode it. Thereby, the CR DUs first decode the common message and remove its interference via applying SIC, and then decode their intended private messages.  Based on the aforementioned RS and decoding strategy,  in what follows, the E2E sum-rate expressions of both hops  are provided.}

%Based on the aforementioned discussion, we provide the expression of E2E sum-rate as follows. 
In the first hop, the transmitted signals from the CH DUs of the $\mathcal{S}_{j,1}$ and the $\mathcal{S}_{j,2}$ D2D links are expressed as, $x_{1}^{S-R}=\sum_{i=1}^2 \sqrt{P_{j,l,n}^{(1,i)}}s_{j,i}^{(1)} $ and $x_{2}^{S-R}=\sqrt{P_{j,l,n}^{(2)}}s_{j}^{(2)} $, respectively. Here, $P_{j,l,n}^{(1,i)}$ is the transmit power for the $i$-th data stream ($i=1,2$) of the CH DU in the $\mathcal{S}_{j,1}$ D2D link, and $P_{j,l,n}^{(2)}$ is the transmit power of the CH DU in the $\mathcal{S}_{j,2}$ D2D link.  Let $h_{j,l,n}^{(1)}$  and $h_{j,l,n}^{(2)}$ be the channel gains from the CH DUs of   the $\mathcal{S}_{j,1}$ and the $\mathcal{S}_{j,2}$ D2D links to the $l$-th eRRH over the $n$-th RRB,  respectively, and $h_{j,l,n}^{(1)}>h_{j,l,n}^{(2)}$.  Hence, the received signal at the $l$-th eRRH over the $n$-th RRB is obtained as 
$Y_{l,n}=h_{j,l,n}^{(1)}x_{1}^{S-R}+h_{j,l,n}^{(2)}x_{2}^{S-R}+n_a+\sqrt{\mathcal{P}_{c,n}}\tilde{h}_{c,l,n}$. Here,   $n_a$ is the additive white Gaussian noise with variance $\sigma^2$;  $\mathcal{P}_{c,n}$ is the uplink transmit power of the $c$-th CUE over the $n$-th RRB; and $\tilde{h}_{c,l,n}$ is the interference channel gain between the $c$-the CUE and the $l$-th eRRH over the $n$-th RRB. The  sum transmission  rate from both CH DUs in the $j$-th device cluster is obtained as $\mathrm{R}_{j,l,n}^{(U)}=R_{j,l,n}^{(1,1)}+R_{j,l,n}^{(2)}+R_{j,l,n}^{(1,2)}$. Here, $R_{j,l,n}^{(1,1)}=\frac{1}{2}\log_2\left(1+\frac{A_{j,l,n}^{(1)}}{B_{j,l,n}^{(1)}}\right)$, $R_{j,l,n}^{(2)}=\frac{1}{2} \log_2\left(1+\frac{A_{j,l,n}^{(2)}}{B_{j,l,n}^{(2)}}\right)$, $R_{j,l,n}^{(1,2)}=\frac{1}{2}\log_2\left(1+\frac{ P_{j,l,n}^{(1,2)} |h_{j,l,n}^{(1)}|^2}{\mathcal{P}_{c,n}|\tilde{h}_{c,l,n}|^2+\sigma^2}\right)$, and  
\begin{subequations}
	\begin{align}
	& A_{j,l,n}^{(1)}=P_{j,l,n}^{(1,1)} |h_{j,l,n}^{(1)}|^2, \hspace{0.2cm}   A_{j,l,n}^{(2)}=P_{j,l,n}^{(2)}|h_{j,l,n}^{(2)}|^2, \label{A}  \\
	& B_{j,l,n}^{(1)}=P_{j,l,n}^{(1,2)}|h_{j,l,n}^{(1)}|^2+P_{j,l,n}^{(2)}|h_{j,l,n}^{(2)}|^2+\mathcal{P}_{c,n}|\tilde{h}_{c,l,n}|^2+\sigma^2,\label{B_1}\\
	&B_{j,l,n}^{(2)}=P_{j,l,n}^{(1,2)}|h_{j,l,n}^{(1)}|^2+\mathcal{P}_{c,n}|\tilde{h}_{c,l,n}|^2+\sigma^2.\label{B_2}
	\end{align}
\end{subequations}

In the second hop, the transmitted signal from the $l$-th eRRH is expressed as $x_{l}^{R-D}=\sqrt{Q_{l,\bm{C},n}}s_c+\sqrt{Q_{l,j,n}^{(1)}}s_{j,p}^{(1)}+\sqrt{Q_{l,j,n}^{(2)}}s_{j,p}^{(2)}$ where $s_c$ denotes the common message encoded for both  CR DUs, and $s_{j,p}^{(1)}$ and $s_{j,p}^{(2)}$ denote the private messages encoded  for the CR DUs of  the $\mathcal{S}_{j,1}$ and the $\mathcal{S}_{j,2}$ D2D links, respectively. Evidently, the $l$-th eRRH broadcasts the common message with $Q_{l,\bm{C},n}$ transmit power, and the private messages with $Q_{l,j,n}^{(1)}$ and $Q_{l,j,n}^{(2)}$ transmit powers. The received signals at the CR DUs of the $\mathcal{S}_{j,1}$ and $\mathcal{S}_{j,2}$  D2D links are expressed as $Y_{j,l,n}^{(D,1)}=g_{j,l,n}^{(1)}x_{l}^{R-D}+\sqrt{\mathcal{P}_{c,n}}\tilde{g}_{c,j,n}^{(1)}+n_a$ and $Y_{j,l,n}^{(D,2)}=g_{j,l,n}^{(2)}x_{l}^{R-D}+\sqrt{\mathcal{P}_{c,n}}\tilde{g}_{c,j,n}^{(2)}+n_a$, respectively.  Here, $g_{j,l,n}^{(1)}$ (resp. $g_{j,l,n}^{(2)}$) is the channel gain between the $l$-th eRRH and CR DU of the $\mathcal{S}_{j,1}$ (resp. $\mathcal{S}_{j,2}$) D2D link over the $n$-th RRB;  and $\{\tilde{g}_{c,j,n}^{(1)}, \tilde{g}_{c,j,n}^{(2)}\}$ denote the gain of the interference channels from the $c$-the CUE over the $n$-th RRB. To determine the common message rate, we denote $\kappa=\arg \min \left(\frac{|g_{j,l,n}^{(1)}|^2}{\mathcal{P}_{c,n}|\tilde{g}_{c,j,n}^{(1)}|^2+\sigma^2},\frac{|g_{j,l,n}^{(2)}|^2}{\mathcal{P}_{c,n}|\tilde{g}_{c,j,n}^{(2)}|^2+\sigma^2}\right)$.
The data rate scheduled for the common message is expressed as $\tilde{R}_{l,j, n}^{(\bm{C})}=\frac{1}{2} \log_2\left(1+\frac{\widehat{A}_{j,l,n}^{(1)}}{\widehat{B}_{j,l,n}^{(1)}}\right)$ where
\begin{equation}
\label{CM_signal}
\widehat{A}_{j,l,n}^{(1)}=Q_{l,\bm{C},n}|g_{j,l,n}^{(\kappa)}|^2 \quad\text{and} \quad
\widehat{B}_{j,l,n}^{(1)}= \left(Q_{l,j,n}^{(1)}+Q_{l,j,n}^{(2)}\right)|g_{j,l,n}^{(\kappa)}|^2+\mathcal{P}_{c,n}|\tilde{g}_{c,j,n}^{(\kappa)}|^2+\sigma^2.
\end{equation} 
In \eqref{CM_signal}, $\kappa \in \{1,2\}$. After removing the common message's interference, each CR DU decodes its own private message while considering  the other CR DU's private message as noise. Thus, the scheduled data rates for the private messages of the CR DUs in the $\mathcal{S}_{j,1}$ and $\mathcal{S}_{j,2}$ D2D links are obtained as $\tilde{R}_{l,j,n}^{(1)}=\frac{1}{2}\log_2\left(1+\frac{\widehat{A}_{j,l,n}^{(2)}}{\widehat{B}_{j,l,n}^{(2)}}\right)$ and $\tilde{R}_{l,j,n}^{(2)}=\frac{1}{2}\log_2\left(1+\frac{\widehat{A}_{j,l,n}^{(3)}}{\widehat{B}_{j,l,n}^{(3)}}\right)$, respectively. Here, 
\begin{subequations}
	\begin{align}
	&\widehat{A}_{j,l,n}^{(2)}=Q_{l,j,n}^{(1)}|g_{j,l,n}^{(1)}|^2, \hspace{0.2cm}  \widehat{A}_{j,l,n}^{(3)}=Q_{l,j,n}^{(2)}|g_{j,l,n}^{(2)}|^2\label{A_2nd_hop_2}\\
	&\widehat{B}_{j,l,n}^{(2)}= Q_{l,j,n}^{(2)}|g_{j,l,n}^{(1)}|^2+\mathcal{P}_{c,n}|\tilde{g}_{c,j,n}^{(1)}|^2+\sigma^2\\
	& 	\widehat{B}_{j,l,n}^{(3)}= Q_{l,j,n}^{(1)}|g_{j,l,n}^{(2)}|^2+\mathcal{P}_{c,n}|\tilde{g}_{c,j,n}^{(2)}|^2+\sigma^2.\label{A_2nd_hop_3}
	\end{align}
\end{subequations}
The received sum data rate of the CR DUs in the $j$-th device-cluster is obtained as $\mathrm{R}_{j,l,n}^{(D)}=\tilde{R}_{l,j, n}^{(\bm{C})}+\tilde{R}_{l,j, n}^{(1)}+\tilde{R}_{l,j, n}^{(2)}$. Therefore, the E2E sum-rate of the $j$-th device cluster over the  $l$-th  eRRH and the $n$-th RRB is obtained as $R_{j,l,n}^{e2e}=\min\left(\mathrm{R}_{j,l,n}^{(U)},\mathrm{R}_{j,l,n}^{(D)}\right)$. 

\vspace{-0.4cm}
\subsection{Problem Formulation}
We denote by $\mathcal{S}=\left\{\mathcal{S}_1,\mathcal{S}_2, \cdots, \mathcal{S}_{\left|\mathcal{T}\right|}\right\}$  the collection of all the device-clusters where $\mathcal{S}_j$ is the set of  D2D links belong to the $j$-th device cluster and $\mathcal{T}=\{1,2, \cdots, \lceil\frac{M}{2}\rceil\}$ is  the set of indices of the device-clusters.  
Assuming that both the $l$-th eRRH and the $n$-th RRB are allocated to the $j$-th device-cluster, the CH DUs' and the eRRH's transmit power vectors in the $j$-th device-cluster are denoted by $\mathbf{P}_{j,l,n}=\left[P_{j,l,n}^{(1,2)}, P_{j,l,n}^{(1,2)}, P_{j,l,n}^{(2)}\right]^{T}$ and $\mathbf{Q}_{j,l,n}=\left[Q_{l,j,n}^{(1)}, Q_{l,j,n}^{(2)}, Q_{l,\bm{C},n}\right]^T$, respectively. The overall transmit power vectors are denoted by $\mathbf{P}=\left[\mathbf{P}_{j,l,n}\right]_{j \in \mathcal{T}, l \in  \mathcal{L}, n \in \mathcal{N}}$ and $\mathbf{Q}=\left[\mathbf{Q}_{j,l,n}\right]_{j \in \mathcal{T}, l \in  \mathcal{L}, n \in \mathcal{N}}$.   We also introduce two binary variables, $x_{l,j} \in \{0,1\}$ and $y_{n,l,j} \in \{0,1\}$ such that $x_{l,j}=1$ if the $l$-th eRRH is selected to assist the $j$-th device-cluster and $x_{l,j}=0$ otherwise; and $y_{n,l,j}=1$ if the $n$-th RRB is assigned to the $j$-th device-cluster and $l$-th eRRH and $y_{n,l,j}=0$ otherwise.  \textcolor{black}{In the proposed system model,  an increase of the transmit power of the D2D links  affects the uplink F-RAN through the increased interference. To resolve such a conflict, we aim to maximize the E2E sum-rate of the device-clusters while using less transmit power. To this end, we consider maximizing the following function for each device-cluster.}
\begin{equation}
\label{Optimization_SOOP}
\begin{split}
\mathrm{U}_j=  \sum_{l=1}^L\sum_{n=1}^N x_{l,j}y_{n,l,j}\left(R_{j,l,n}^{e2e}-\mu_n
\left(\sum_{i=1}^3 \mathbf{P}_{j,l,n}[i]+\sum_{i=1}^3\mathbf{Q}_{j,l,n}[i] \right)\right)\\
\end{split}
\end{equation}
where $\mu_n \geq 0, \forall n$ is a weight factor. Note that, eq. \eqref{Optimization_SOOP} is the overall utility function of the $j$-th device-cluster. Particularly, the second term of \eqref{Optimization_SOOP}  is interpreted as the  cost charged to the $j$-th device-cluster for re-using the $n$-th RRB, and it is also proportional to the uplink interference at the CBS caused by the $j$-th device-cluster  over the $n$-th RRB. Thus,  the weight factor, $\mu_n$, represents the $n$-th RRB's price, $\forall n \in \mathcal{N}_{sc}$. Since each RRB is shared between a device-cluster and an uplink CUE, CBS needs to control prices of the RRBs to satisfy the predefined uplink interference constraint. Hence, we consider $\{\mu_n\}$ as  the optimization variables in our problem formulation and  the  suitable values of $\{\mu_n\}$ are determined from the proposed solution. 
\setlength{\textfloatsep}{0pt}
\begin{table*}
	%	\vspace*{-0.6cm}
	\begin{normalsize} 
		\begin{equation}
		\label{Optimization_Main}
		\begin{split}
		&\text{P0}:\hspace{0.1cm}\max_{\substack{\mathcal{S}, \mathbf{P} \geqq \bm{0}, \mathbf{Q} \geqq \bm{0}, \mathbf{x}\in \{0,1\}, \mathbf{y}\in \{0,1\}, \bm{\mu} \geqq 0 }} \sum_{j \in \mathcal{T}} \mathrm{U}_j \\
		&\text{s.t.} \begin{cases}
		& \hspace{-0.3cm}\text{C1:} \hspace{0.2 cm} \left|\mathcal{S}_j\right| \leq 2, S_j \cap S_{j'}= \varnothing,  \cup _{j \in \mathcal{T}} \mathcal{S}_j=\mathcal{M}; \forall j \neq j',  j,j' \in \mathcal{T}\\
		& \hspace{-0.3cm}\text{C2:} \hspace{0.2cm} \sum_{l=1}^L x_{l,j}=1; \sum_{j \in \mathcal{T}} x_{l,j} \leq N_{R}, \forall l \in \mathcal{L}, j \in \mathcal{T}\\
		& \hspace{-0.3cm}\text{C3:}  \hspace{0.2cm} \sum_{j \in \mathcal{T}}  \sum_{l=1}^L y_{n,l,j}=1; \sum_{n=1}^N \sum_{l=1}^L y_{n,l,j} \geq 1; y_{n,l,j} \leq x_{l,j},  \forall  n \in \mathcal{N}_{sc},l \in \mathcal{L}, j \in \mathcal{T}\\
		& \hspace{-0.3cm}\text{C4:}  \hspace{0.2 cm}\max\left( \sum_{l=1}^L\sum_{n=1}^N\sum_{i=1}^2 x_{l,j}y_{n,l,j} \mathbf{P}_{j,l,n}[i], \sum_{l=1}^L\sum_{n=1}^N x_{l,j}y_{n,l,j} \mathbf{P}_{j,l,n}[3]\right) \leq P^{(D)}_{\mbox{max}}, \\
		%& \quad\quad \sum_{l=1}^L\sum_{n=1}^N x_{l,j}y_{n,l,j} \mathbf{P}_{j,l,n}[3] \leq P^{(D)}_{\mbox{max}}, \\
		& \quad\quad \sum_{l=1}^L\sum_{n=1}^N\sum_{i=1}^3 x_{l,j}y_{n,l,j} \mathbf{Q}_{j,l,n}[i] \leq P^{(R)}_{\mbox{max}}/N_R, \forall j \in  \mathcal{T}\\
		&\hspace{-0.3cm}\text{C5:} \hspace{0.2cm} \sum_{j \in \mathcal{T}}\sum_{l=1}^L x_{l,j}y_{n,l,j} \max\left(I_{j,l,n}^{(1)}, I_{j,l,n}^{(2)}\right) \leq I_{th}, \forall n \in \mathcal{N}_{sc}.\\
		\end{cases} 
		\end{split}
		\end{equation}
	\end{normalsize}
	%	\vspace*{-1.1cm}
	\hrulefill
\end{table*}

The resource optimization problem is formulated as  $\text{P0}$ at the top of next page. In $\text{P0}$,  $\text{C1}$ is a device-cluster formation constraint implying that  device-clusters are pairwise disjoint and consist of two D2D links; $\text{C2}$ implies that each device-cluster is assisted by only one eRRH whereas a given eRRH  can assist maximum $N_R$ device-clusters; $\text{C3}$ implies the orthogonal allocation of the RRBs among the device-clusters; $\text{C4}$ provides the transmit power allocation constraints in each device-cluster with $P^{(D)}_{\mbox{max}}$ and $P^{(R)}_{\mbox{max}}$ as the maximum transmit power limit of the CH DU and eRRH, respectively;  and $\text{C5}$ implies  that for each shared RRB, the uplink interference at the CBS caused by the device-cluster(s) will be bounded. In the constraint $\text{C5}$,  $I_{j,l,n}^{(1)}$ and $I_{j,l,n}^{(2)}$ denote the uplink interference caused  by the CH DUs and the associated eRRH of the $j$-th device-cluster over the  $n$-th RRB, respectively, and $I_{th}$ is the acceptable  interference threshold at the CBS.

\begin{lem}
	\label{Lemma_1}
	$\text{P0}$ is an NP-hard optimization problem.
\end{lem}

\proof The proof is provided in Appendix A.

\textcolor{blue}{Since $\text{P0}$ is NP-hard,  in the following two sections, we obtain a sub-optimal yet efficient solution to $\text{P0}$ by decoupling device-clustering and resource allocation in two sub-problems. In consistent with the existing literature \cite{ML_Clustering,ML_clustering_2},  we use the unsupervised-learning and optimization based algorithms, respectively, to solve the  device-clustering and resource-allocation sub-problems.}

\vspace{-0.4cm}
\section{Proposed Solution: Device-clustering}
\textcolor{black}{The essence of device-clustering is to pair two D2D links so that the channel disparity between both CH DUs and  the channel disparity between both  CR DUs are reduced.  For simplicity, we define the channel disparity between the clustered CH DUs (and the clustered  CR DUs)  as the Euclidean distance between  their channel  gain matrices.  By linearly combining such distance metrics,  we obtain the overall channel disparity between the clustered D2D links. 
Note that the reduced channel disparity between the clustered D2D links implies the reduced  channel strength difference between both CH DUs and both CR DUs of  the  device-cluster. 
In a given RRB, the reduced channel strength difference between both CH DUs can improve the transmission rate of the weak CH DU in the first hop, and  the reduced channel strength difference between both CR DUs can improve the common message rate  in the second hop.  Hence,  the reduced channel disparity between the paired D2D links can improve the E2E sum-rate of the device-cluster(s). However, to pair D2D links of low channel disparity, CBS requires the entire channel matrix of all the D2D links in the system.  In practice, dimension of a D2D link's entire channel matrix can be large, and uploading such a large channel matrix for all the D2D links at CBS  requires prohibitive training overhead, especially in a large-scale network.
	%Moreover, an exhaustive-search based clustering while requiring no channel information is infeasible for the practical network. 
	To overcome such a challenge, we desire a low-complexity device-clustering algorithm while reducing the required  features to obtain the device-clusters. In the literature, PCA is shown to be a suitable  technique to effectively reduce  dimension of the features for clustering \cite{Mahdi_Rev_1}.  In the considered system setting, 2D-PCA  extracts the information of channel matrix by using a small number of uncorrelated vectors, namely, the principle component vectors (PCVs).  These PCVs contain \textit{most of the variability}  of the channel matrix of a D2D link,  and as a result, such PCVs provide the suitable features for clustering the D2D links. Consequently, CBS  requires only a set of PCVs for each D2D link.  Such a dimension-reduction technique relieves  the requirement of uploading the entire channel matrices to the CBS.  Essentially, by applying 2D-PCA technique, the required  training overhead for device-clustering process can be reduced substantially.} In the ensuing sub-section, we  develop  a 2D-PCA enabled device-clustering algorithm. 

\vspace{-0.41cm}
\subsection{ Device-clustering Algorithm Using 2D-PCA} 
%The required steps for clustering devices are discussed as follows. 
%\vspace{-0.5cm}
\subsubsection{Feature selection} 
\textcolor{black}{We emphasize   that clustering of the communication links belongs to the unsupervised learning, and the channel gain matrices of the links can be cast as the feature set of such an unsupervised learning  method \cite{ML_Clustering}. To this end, the channel gain matrices of both CH DU and CR DU are cast as the two features for each D2D link.}  Let $\mathbb{H}_m\in \mathbb{R}^{L \times N}$  and $\mathbb{G}_m\in \mathbb{R}^{L \times N}$ be the entire channel gain matrices of the CH DU and CR DU of the $m$-th D2D link, respectively, where $\mathbb{H}_m=\left[\mathbf{H}_{m,1}, \mathbf{H}_{m,2}, \cdots, \mathbf{H}_{m,L} \right]^T$ and $\mathbb{G}_m=\left[\mathbf{G}_{m,1}, \mathbf{G}_{m,2}, \cdots, \mathbf{G}_{m,L} \right]^T$.  Here, both  $\mathbf{H}_{m,l}$ and $\mathbf{G}_{m,l}$ are two $N$-dimensional vectors where each element of $\mathbf{H}_{m,l}$ (resp. $\mathbf{G}_{m,l}$) represents the channel gain between the CH DU (resp. CR DU) and the $l$-th eRRH over an orthogonal RRB. Thus, $\{\mathbb{H}_m, \mathbb{G}_m\}_{ m \in \mathcal{M}}$ provides the entire feature set for clustering.

\subsubsection{Feature transformation} 
The next step is to reduce the dimension of the feature set. To do so, we transform both $\mathbb{H}_m$ and $ \mathbb{G}_m$, $\forall m \in \mathcal{M}$, to a lower-dimension matrix by using an orthogonal projection as such the maximum amount of variation is retained. In what follows, we briefly present the key steps, and  the detailed analysis can be found in \cite{PCA_Basic_1_Rev_1}. 
The optimal projection directions are determined as the eigen-vectors of the auto-covariance (ACV) matrices of $\mathbb{H}_m$ and $\mathbb{G}_m$,  defined as, $\mathbb{S}_m \triangleq \mathbb{E}\left[\left(\mathbb{H}_m-\mathbb{E}[\mathbb{H}_m]\right)^T\left(\mathbb{H}_m-\mathbb{E}[\mathbb{H}_m]\right)\right]$ and $\mathbb{K}_m \triangleq \mathbb{E}\left[\left(\mathbb{G}_m-\mathbb{E}[\mathbb{G}_m]\right)^T\left(\mathbb{G}_m-\mathbb{E}[\mathbb{G}_m]\right)\right]$, respectively. In the absence of channel statistics information, the ACV matrices can be computed from the channel training samples. Suppose, $\left\{\mathbb{H}_m^{(i)}\right\}$ and  $\left\{\mathbb{G}_m^{(i)}\right\}$, $i=1,2,\cdots, I$, are the $I$ channel training samples for the CH DU and CR DU of the $m$-th D2D link, respectively, and $\overline{\mathbb{H}_m}$ and $\overline{\mathbb{G}_m}$ are the  corresponding sample averages. Therefore, the ACV matrices are evaluated as $\mathbb{S}_m= \frac{1}{I} \sum_{i=1}^{I}\left(\mathbb{H}_m^{(i)}-\overline{\mathbb{H}_m}\right)^T\left(\mathbb{H}_m^{(i)}-\overline{\mathbb{H}_m}\right)$ and $\mathbb{K}_m= \frac{1}{I} \sum_{i=1}^{I}\left(\mathbb{G}_m^{(i)}-\overline{\mathbb{G}_m}\right)^T\left(\mathbb{G}_m^{(i)}-\overline{\mathbb{G}_m}\right)$.  Since both $\mathbb{S}_m$ and   $\mathbb{K}_m$ are symmetric and $N\times N$ square matrices,  the singular vale decomposition (SVD) of $\mathbb{S}_m$ and  $\mathbb{K}_m$ is expressed as $\mathbb{S}_m=\mathbb{V}_m \mathbf{\Sigma}_m \mathbb{V}_m^{T}$ and $\mathbb{K}_m=\mathbb{W}_m \mathbf{\Delta}_m \mathbb{W}_m^{T}$ . Here, $\mathbb{V}_m$ (resp. $\mathbb{W}_m$) is a matrix of $N$ orthogonal eigen-vectors of $\mathbb{S}_m$ (resp.  $\mathbb{K}_m$  ) and it is sorted in a decreasing order of the eigen-values, and $\mathbf{\Sigma}_m$ (resp.  $\mathbf{\Delta}_m$ ) is an $N \times N$ diagonal matrix containing squares of the eigen-values of $\mathbb{S}_m$  (resp.  $\mathbb{K}_m$ ). A set of the first $d$ eigen-vectors, corresponding to the first largest $d$ eigen-values of  $\mathbb{S}_m$ and $\mathbb{K}_m$, constitute the optimal projection directions. For the CH DU and CR DU of the  $m$-th D2D link, the first $d$  PCVs  are computed as $\mathbf{Y}_{m,k}=\mathbb{H}_m \mathbf{V}_{m}^{(k)}$ and $\mathbf{Z}_{m,k}=\mathbb{G}_m \mathbf{W}_{m}^{(k)}$,  respectively, $\forall k=1,2,\cdots, d$. Here, $\mathbf{V}_{m}^{(k)}$ and $\mathbf{W}_{m}^{(k)}$ are the $k$-th column vector of  $\mathbb{V}_m$ and $\mathbb{W}_m$, respectively. \textcolor{black}{Essentially, we transform the original feature set, $\{\mathbb{H}_m, \mathbb{G}_m\}$ to a \textit{modified} feature set of the reduced dimension, $\{\mathbb{B}_m^{(1)}, \mathbb{B}_m^{(2)}\}$, where   $\mathbb{B}_m^{(1)}=[\mathbf{Y}_{m,1}, \mathbf{Y}_{m,2}, \cdots, \mathbf{Y}_{m,d}] \in \mathbb{R}^{L \times d}$ and $\mathbb{B}_m^{(2)}=[\mathbf{Z}_{m,1}, \mathbf{Z}_{m,2}, \cdots, \mathbf{Z}_{m,d}] \in \mathbb{R}^{L \times d}$.  Both $\mathbb{B}_m^{(1)}$ and $\mathbb{B}_m^{(2)}$ provide the suitable features for clustering the D2D links,  and we denote them  by feature matrices \cite{PCA_Basic_1_Rev_1}. }%The required number of PCVs to construct feature matrices is a system defined parameter.}

\subsubsection{Algorithm development}
\textcolor{black}{The Euclidean distance between two different feature matrices provides a  viable metric to decide whether the corresponding objects (i.e., the D2D links) belong to the same or different clusters \cite{PCA_Basic_1_Rev_1}.  In particular,  the Euclidean distance between the feature matrices of two D2D links of the same and different clusters are small and  large, respectively. Accordingly, we adopt the Euclidean distance between feature matrices  as a function for clustering the D2D links. The Euclidean distance   between  feature matrices  of the CH DUs of the $m$-th and the $k$-th D2D links is denoted by $\phi(\mathbb{B}_m^{(1)} , \mathbb{B}_m^{(1)} ) \triangleq\left|\left|\mathbb{B}_m^{(1)}-\mathbb{B}_k^{(1)}\right|\right|_2$.  Similarly,  the Euclidean distance   between the feature matrices of the CR DUs of the $m$-th and the $k$-th D2D links is denoted by $\phi(\mathbb{B}_m^{(2)} , \mathbb{B}_m^{(2)} ) \triangleq\left|\left|\mathbb{B}_m^{(2)}-\mathbb{B}_k^{(2)}\right|\right|_2$.  We define $d(\mathbb{B}_m , \mathbb{B}_k )=w \phi(\mathbb{B}_m^{(2)} , \mathbb{B}_m^{(2)} )+(1-w)\phi(\mathbb{B}_m^{(1)} , \mathbb{B}_m^{(1)} )$,  where $w \in (0,1)$ is a weight factor.  Particularly, by setting  $w=1$, we obtain a measure of the channel disparity between  the CH DUs of two D2D links and  by setting  $w=0$, we obtain   a measure of the channel disparity between the  CR DUs of two D2D links.  Hence, $d(\mathbb{B}_m , \mathbb{B}_k )$ provides a measure of  the overall channel disparity between the $m$-th and the $k$-th D2D links. In fact, the small and large values of $d(\mathbb{B}_m , \mathbb{B}_k )$ imply the low and high channel disparity between the $m$-th and the $k$-th D2D links, respectively. Recall,  our goal is to reduce the channel disparity between the clustered D2D links.  Consequently,  the device-clustering is cast as an optimization problem of pairing the  D2D links  so that the maximum distance between the feature matrices of  the clustered D2D links is minimized. Such an optimization problem is formulated as}
\begin{subequations}
	\begin{align}
	&\text{P1}:\hspace{0.2cm} \min_{ \mathcal{S}_1, \mathcal{S}_2, \cdots, \mathcal{S}_{|\mathcal{T}|}} \max_{j \in \mathcal{T}} \mathtt{d}(\mathbb{B}_{\mathcal{S}_{j,1}}, \mathbb{B}_{\mathcal{S}_{j,2}}) \\
	&\text{s.t.}\begin{cases}
	&\mathcal{S}_j=\{\mathcal{S}_{j,1}, \mathcal{S}_{j,2}\},  \mathcal{S}_j \cap \mathcal{S}_{j' \neq j}= \varnothing,  \forall j, j' \in \mathcal{T}\\  
	&\mathcal{S}_{j,1} \in \left\{1,2, \cdots, M\right\}, \mathcal{S}_{j,2} \in \left\{1,2, \cdots, M\right\}, \mathcal{S}_{j,1} \neq \mathcal{S}_{j,2}, \forall j \in  \mathcal{T}.  \\
	\end{cases} 
	\end{align}
\end{subequations}
\textcolor{black}{The $m$-th and the $k$-th D2D  links will  be in the same device-cluster if the channel disparity between the $m$-th and the $k$-th D2D links is small. Hence, the $m$-th D2D link is preferred to form a device-cluster with the $k$-th D2D link over the $e$-th D2D link when  $\mathtt{d}(\mathbb{B}_m, \mathbb{B}_k) < \mathtt{d}(\mathbb{B}_m, \mathbb{B}_e)$ is satisfied.} Thus, the quantity $\mathtt{d}(\mathbb{B}_m, \mathbb{B}_e)-\mathtt{d}(\mathbb{B}_m, \mathbb{B}_k)$ can be interpreted as the \textit{loss function} for clustering the $m$-th D2D link with its less preferred D2D link. Therefore, $\text{P1}$ can be  solved heuristically by minimizing such a loss function for clustering any two D2D links. We propose a heuristic algorithm, namely, Algorithm \ref{Algorithm1} to solve  $\text{P1}$.  In Algorithm \ref{Algorithm1}, $\mathcal{F}$ denotes the set of the clustered D2D links, $\mathcal{F}^{\complement}$ denotes the set of un-clustered D2D links, and $\Delta_m$ is a \textit{cost function} that implies the cost experienced by the $m$-th D2D link for not being clustered with its most preferred D2D link. Algorithm \ref{Algorithm1} first selects a D2D link from the $\mathcal{F}^{\complement}$ set having the maximum cost function, and pairs it with its most preferred D2D link. Subsequently, Algorithm \ref{Algorithm1} removes these two D2D links from the $\mathcal{F}^{\complement}$ set. These two steps are  repeated iteratively until all the D2D links are clustered. 
\setlength{\textfloatsep}{0pt}
\begin{algorithm}
	\caption{Low-complexity device-clustering algorithm}
	\label{Algorithm1}
	\begin{algorithmic}[1]
		\State \textbf{Input:} Feature matrices, $\{\mathbb{B}_m^{(1)}, \mathbb{B}_m^{(2)}\}$, $\forall m \in \mathcal{M}$; weight factor $w \in (0,1)$.
		
		\State \textbf{Initialize:} $\mathcal{F}=\varnothing$, $\mathcal{F}^{\complement}=\mathcal{M}$,  	$ \left\{\mathcal{S}_1,\mathcal{S}_2, \cdots, \mathcal{S}_{|\mathcal{T}|}\right\}=\varnothing$, $j=1$.
		\While{$\mathcal{F}^{\complement} \neq \varnothing$}
		
		\State Determine $\hat{k}(m)=\arg \min_{k \in \mathcal{F}^{\complement}, k \neq m} \mathtt{d}(\mathbb{B}_m, \mathbb{B}_k), \forall m \in \mathcal{F}^{\complement} $.
		
		\State Calculate  $\Delta_m=\min_{k \in \mathcal{F}^{\complement}, k \neq \{m,\hat{k}(m)\}}\mathtt{d}(\mathbb{B}_m, \mathbb{B}_k)-\mathtt{d}(\mathbb{B}_m, \mathbb{B}_{\hat{k}(m)}),  \forall m \in \mathcal{F}^{\complement} $.
		
		\State Select $m^*= \arg \max_{ m \in \mathcal{F}^{\complement}} \Delta_m $.
		
		\State $\mathcal{S}_j \leftarrow \mathcal{S}_j \bigcup \{m^*, \hat{k}(m^*)\}$, $j=j+1$;
		
		\State Update $\mathcal{F} \leftarrow \mathcal{F} \bigcup \{m^*, \hat{k}(m^*)\}$ and $\mathcal{F}^{\complement} \leftarrow \mathcal{M} \setminus \mathcal{F} $.
		
		\EndWhile
		
		\For{$j=1:|\mathcal{T}|$, $j'=1:|\mathcal{T}|, \quad \text{and}\quad j' \neq j$}

		\State If  $\mathtt{d}(\mathbb{B}_{\mathcal{S}_{j,1}}, \mathbb{B}_{\mathcal{S}_{j',2}})+\mathtt{d}(\mathbb{B}_{\mathcal{S}_{j',1}}, \mathbb{B}_{\mathcal{S}_{j,2}}) < \mathtt{d}(\mathbb{B}_{\mathcal{S}_{j,1}}, \mathbb{B}_{\mathcal{S}_{j,2}})+\mathtt{d}(\mathbb{B}_{\mathcal{S}_{j',1}}, \mathbb{B}_{\mathcal{S}_{j',2}})$, swap the current members between the $j$-th  and $j'$-th device-clusters.
		
		\EndFor
		%	\EndFor
		
		\State \textbf{Output:} Non-overlapping device-clusters $\mathcal{S}_1, \mathcal{S}_2, \cdots, \mathcal{S}_{|\mathcal{T}|}$.
	\end{algorithmic}
\end{algorithm}

\vspace*{-0.4cm}
\subsection{Properties of Algorithm \ref{Algorithm1} }
\subsubsection{Complexity and implementation}
\textcolor{black}{The computational complexity of Algorithm \ref{Algorithm1} depends on the complexity of computing the feature matrices and determining the set of device-clusters. The complexity of SVD decomposition of an $N \times N$ matrix is $\mathcal{O}\left(N^3\right)$.  Thus, the overall complexity of computing the feature matrices for all the D2D links is $\mathcal{O}\left(MN^3\right)$. Meanwhile, at each iteration of  Steps 3-9, Algorithm \ref{Algorithm1} requires $|\mathcal{F}^{\complement}|$  computations to form a device-cluster.  Since at each iteration of Algorithm \ref{Algorithm1}, two D2D links are clustered together, $|\mathcal{F}^{\complement}|$ evolves as $M, M-2,  M-4,\cdots $ as the number of iterations is increased. Therefore,  the required computations for executing Steps 3-9 of Algorithm \ref{Algorithm1} are $\lceil\frac{M}{2}\rceil^2$ and $M(M+2)/4$ for odd and even $M$, respectively,  approximated as, $\mathcal{O}\left(M^2\right)$. Thus, the overall computational complexity of Algorithm \ref{Algorithm1} is 
	$\mathcal{O}\left(MN^3+M^2\right)$.  Using the following procedures, Algorithm 1 can be implemented in a practical network. In offline (i.e., before the communication takes place), both  CH DU and CR DU of each D2D link estimate the ACV matrix of their channel and compute SVD decomposition of the estimated ACV matrix. In online, both CH DU and CR DU  of each D2D link  first compute the predefined number of PCVs  and upload these PCVs to their nearest eRRH. The eRRHs forwards the received PCVs to the CBS.  Finally, CBS determines the device-clusters by executing Algorithm 1.}

\subsubsection{Pareto-efficiency}  
The output of Algorithm \ref{Algorithm1} will be \textit{Pareto-efficient} if it does not contain any \textit{Pareto-improvement pair}. We  define the notion of \textit{Pareto-improvement pair} as follows. Two different device-clusters  $\mathcal{S}_j=\{m,k\}$ and $\mathcal{S}_j'=\{m',k'\}$ will be \textit{Pareto-improvement pair} if  by swapping their current members, their overall \textit{channel disparity} is strictly reduced, i.e.,  the condition $\mathtt{d}(\mathbb{B}_m, \mathbb{B}_{k'})+\mathtt{d}(\mathbb{B}_{m'}, \mathbb{B}_k) < \mathtt{d}(\mathbb{B}_m, \mathbb{B}_k)+\mathtt{d}(\mathbb{B}_{m'}, \mathbb{B}_{k'})$ is satisfied. 
%The device-cluster set  $\mathcal{S}$  is \textit{Pareto-efficient} if any $\mathcal{S}'$ set does not exist where compared to the device-clusters in $\mathcal{S}$ set,  at least one device-cluster in $\mathcal{S}'$ set has a strictly smaller feature distance between its component D2D links, and rest of the device-clusters in $\mathcal{S'}$ set have similar feature distances between their component D2D links.

\begin{proposition}
	\label{Proposition 1}
	Algorithm \ref{Algorithm1} provides a Pareto-efficient solution to $\text{P1}$.
\end{proposition}

\proof The proof is provided in Appendix B.

\textcolor{blue}{\textit{Remark 2}:   Note that a global optimal solution to $\text{P1}$ is also Pareto-efficient. However, a given Pareto-efficient solution to $\text{P1}$ is not necessarily the global optimal. Therefore,  Algorithm 1 may not obtain a global optimal solution to   $\text{P1}$. Nevertheless, our simulation result shows that the performance gap between the device-clusters obtained by Algorithm 1 and the device-clusters obtained by an exhaustive search is small. Consequently, Algorithm 1 is efficient.}

\vspace*{-0.3cm}
\section{Proposed Solution: Resource Allocation}
We assume that the device-clusters  are obtained from Algorithm \ref{Algorithm1}. Hence, $\text{P0}$ is expressed as
\begin{equation}
\label{Optimization_P2}
\begin{split}
&\text{P2}:\quad \max_{\mathbf{P} \geqq \bm{0}, \mathbf{Q} \geqq \bm{0}, \mathbf{x}\in \{0,1\}, \mathbf{y} \in \{0,1\}, \bm{\mu} \geqq 0 }\sum_{j \in \mathcal{T}} \mathrm{U}_j \quad
\text{s.t.} \quad \text{C2, C3, C4, C5}.
\end{split}
\end{equation}
\textcolor{black}{$\text{P2}$ is NP-hard and consequently, a global optimal solution to $\text{P2}$ is intractable. Note that, in our considered system model,  power allocation requires only local information of the device-clusters whereas the assignment of eRRHs  and RRBs among the device-clusters as well as interference-aware pricing of the RRBs require global network information. Accordingly,  the resource allocation tasks of $\text{P2}$ can be decomposed between CBS and device-clusters. Stackelberg game is well-suited to efficiently solve a bi-level decomposed resource allocation problem \cite{D2D_C-RAN}. Hence, we employ Stackelberg game to solve  $\text{P2}$.} In the proposed  Stackelberg game, the CBS and device-clusters play the role of leader and followers, respectively. The device-clusters solve the following (non-convex) power allocation problem. 
\begin{equation}
\label{F-problem}
\begin{split}
&\text{P2.1}: \quad  \max_{\mathbf{P} \geqq \bm{0}, \mathbf{Q} \geqq \bm{0}} \left\{\sum_{j \in \mathcal{T}} \mathrm{U}_j\bigg|\text{C4}\right\}.\\
\end{split}
\end{equation}
To schedule RRBs and eRRHs among the device-clusters, CBS  solves the following assignment problem.
\begin{equation}
\label{L-problem}
\begin{split}
&\text{P2.2}: \quad  \max_{\mathbf{x}\in \{0,1\}, \mathbf{y} \in \{0,1\}} \left\{\sum_{j \in \mathcal{T}} \mathrm{U}_j\bigg|\text{C2}, \text{C3}\right\}.\\
\end{split}
\end{equation}
Finally, CBS updates the prices of the RRBs by solving the following optimization problem.
\begin{equation}
\label{L-problem-II}
\begin{split}
&\text{P2.3}: \quad  \max_{\bm{\mu} \geqq 0} \left\{\sum_{j \in \mathcal{T}} \mathrm{U}_j\bigg|\text{C5}\right\}.\\
\end{split}
\end{equation}
In what follows, we obtain efficient solutions to $\text{P2.1}$, $\text{P2.2}$, and $\text{P2.3}$. 

\vspace{-0.5cm}
\subsection{Solution to $\text{P2.1}$: Power Allocations for the Device-clusters}
\textcolor{black}{Each device-cluster  is allocated with certain orthogonal RRBs exclusively. In particular,  CBS determines prices of the RRBs to control uplink interference, and the allocation of RRBs among the device-clusters depends on the power allocation strategies of the device-clusters. Apparently, each device-cluster competes with one another to obtain as many as RRBs possible while judiciously choosing  the power allocation strategy to improve its own utility. Such a framework is consistent with a non-cooperative game where the device-clusters  act as rational and selfish players, and they are willing to  maximize their individual payoff.  Accordingly,  the non-cooperative game theory is adopted to solve $\text{P2.1}$.}
\subsubsection{Potential Game Formulation}
%Let the assignment of the eRRH(s) and RRB(s)  among the device-clusters and the RRBs' prices be given. 
Without loss of generality, let the $j$-th device-cluster, $\forall j \in \mathcal{T}$, be assigned with the $l_j$-th eRRH and  the $\mathcal{N}_j$ set of RRBs where $\bigcup_{j=1}^{\left|\mathcal{T}\right|}\mathcal{N}_j=\mathcal{N}_{sc}$, $\mathcal{N}_j \cap \mathcal{N}_j'=\varnothing$, $\forall j\neq j'$, and $\bigcup_{j=1}^{\left|\mathcal{T}\right|}l_j\subset \mathcal{L}$. 
To solve $\text{P2.1}$, a  non-cooperative power-control game (NCPCG) is devised, and it is defined by the tuple $\mathcal{G}=\left\{\mathcal{S}, \left({\Pi}_j\right)_{j \in \mathcal{T} }, \left({\Gamma}_j\right)_{j \in \mathcal{T} }  \right\}$.
Here, $\mathcal{S}$ is the set of players (i.e., non-overlapping device-clusters);  $\Pi_j$ is the strategy space of the $\mathcal{S}_j$ player defined as $\Pi_j =\left\{\mathbf{P}_{j,l_j,n},\mathbf{Q}_{j,l_j,n}\bigg| \mathbf{P}_{j,l_j,n} \geqq \mathbf{0}, \mathbf{Q}_{j,l_j,n} \geqq \mathbf{0}, n \in  \mathcal{N}_j  \right\}$; and $\Gamma_j$ is the payoff of the $\mathcal{S}_j$ player, and it is defined as \eqref{Payoff} at the top of current page.
\setlength{\textfloatsep}{0pt}
\begin{table*}
	%	\vspace*{-0.6cm}
	\begin{normalsize} 
		\begin{equation}
		\label{Payoff}
		\begin{split}
		&\Gamma_j\left(\{\mathbf{P}_{j,l_j,n}, \mathbf{Q}_{j,l_j,n}\} \bigg| \{\bm{x},\bm{y},\bm{\mu}\} \right)=\sum_{n \in \mathcal{N}_j}R_{j,l_j,n}^{e2e}
		-\sum_{n \in \mathcal{N}_j} \mu_n
		\left(\sum_{i=1}^3  \mathbf{P}_{j,l_j,n}[i]+\sum_{i=1}^3\mathbf{Q}_{j,l_j,n}[i] \right)\\
		&-\bm{\sigma_j}^T\left[\sum_{n \in\mathcal{N}_j}\sum_{i=1}^2  \mathbf{P}_{j,l_j,n}[i], \sum_{n \in\mathcal{N}_j} \mathbf{P}_{j,l_j,n}[3], \sum_{n \in\mathcal{N}_j}\sum_{i=1}^3  \mathbf{Q}_{j,l_j,n}[i] \right]^T.
		\end{split}
		\end{equation}	
	\end{normalsize}
	%	\vspace*{-1.1cm}
	\hrulefill
\end{table*}
In \eqref{Payoff}, $\bm{\sigma_j}=[\sigma_{j,1}, \sigma_{j,2}, \sigma_{j,3}]^T$ is a vector of the pricing variables account for the  power consumption cost at the CH DUs and eRRH in the $j$-th device cluster. \textcolor{black}{The aforementioned payoff function enables each device-cluster to  maximize its data rate while   using the transmit power conservatively and causing less interference to the CBS.} A Nash-equilibrium (NE)  of the considered NCPCG is obtained when every player in the game operates by using its best response strategy (BRS). The NE is formally defined as follows. 
\begin{definition}
	\label{NE_Def}
	The tuple $\{\mathbf{P}_{j,l_j,n}^*, \mathbf{Q}_{j,l_j,n}^*\}$ is an NE power allocation strategy of the NCPCG $\mathcal{G}$, if  for all  $\{\mathbf{P}_{j,l_j,n}, \mathbf{Q}_{j,l_j,n} \} \in \Pi_j$, $\mathcal{S}_j \in \mathcal{S}$, the following condition is satisfied.
	\begin{equation}
	\label{NE}
	\Gamma_j\left(\{\mathbf{P}_{j,l_j,n}^*, \mathbf{Q}_{j,l_j,n}^*\} \bigg| \{\bm{x},\bm{y},\bm{\mu}\}\right) \geq \Gamma_j\left(\{\mathbf{P}_{j,l_j,n}, \mathbf{Q}_{j,l_j,n}\} \bigg| \{\bm{x},\bm{y},\bm{\mu}\}\right).
	\end{equation}
\end{definition}
\begin{lem}
	\label{Lemma_2}
	The NCPCG $\mathcal{G}$ is a potential game and it posses an NE power allocation strategy.
\end{lem}
\proof The proof is provided in Appendix C.

\begin{lem}
	\label{Lemma_3}
	The NE power allocation of the NCPCG $\mathcal{G}$ is the BRS of the players. The $j$-th player's BRS, $\forall j \in \mathcal{T}$, is obtained as 
	\begin{equation}
	\label{BRS_I_obj}
	\textbf{BRS}_j: \quad \{\mathbf{P}_{j,l_j,n}^*, \mathbf{Q}_{j,l_j,n}^*\}=\arg\max_{\mathbf{P} \geqq \bm{0}, \mathbf{Q} \geqq \bm{0}} \Gamma_j\left(\{\mathbf{P}_{j,l_j,n}, \mathbf{Q}_{j,l_j,n}\} \bigg| \{\bm{x},\bm{y},\bm{\mu}\} \right)\	\\
	\end{equation}
\end{lem}

The proof of \textit{Lemma} \ref{Lemma_3} is a direct consequence of \cite[Corollary 2.2]{Potential}. In what follows, we obtain NE power allocations for each device-cluster by solving $\textbf{BRS}_j$, $\forall j \in \mathcal{T}$.

\subsubsection{NE  Strategy}
$\textbf{BRS}_j$ is an NP-complete problem with \textit{sum-of-functions-of-ratio} in the objective function.  Capitalizing the auxiliary variable based approach to solve fractional programming problem \cite{W_YU_II}, we establish the following proposition to effectively solve  $\textbf{BRS}_j$.

\begin{proposition}
	$\textbf{BRS}_j$ is equivalent to the following optimization problem:
	\begin{equation}
	\label{BRS_II}
	\begin{split}
	\max_{\mathbf{P} \geqq \bm{0}, \mathbf{Q} \geqq \bm{0}, \mathbf{Z} \geqq \bm{0}, \mathbf{X} \geqq \bm{0}} \mathcal{F}_j^{(1)}\left(\mathbf{Z}, \mathbf{X}\right) + \mathcal{F}_j^{(2)}\left(\mathbf{P}, \mathbf{Q},\mathbf{Z}, \mathbf{X}\right)-\Psi_j\left(\mathbf{P},\mathbf{Q}\right)
	\end{split}
	\end{equation}
	where $\mathbf{Z}=\left[Z_{j,l_j,n}^{(i)}\right]_{i \in \{1,2\}, n \in \mathcal{N}_j}$ and $\mathbf{X}=\left[X_{j,l_j,n}^{(i)}\right]_{i \in \{1,2,3\}, n \in \mathcal{N}_j}$ are the auxiliary variables. Here,   $\mathcal{F}_j^{(1)}\left(\mathbf{Z}, \mathbf{X}\right)$, $\mathcal{F}_j^{(2)}\left(\mathbf{P}, \mathbf{Q},\mathbf{Z}, \mathbf{X}\right)$, and $\Psi\left(\mathbf{P},\mathbf{Q}\right)$  are given by \eqref{F_j_1}, \eqref{F_j_2}, and \eqref{Psi_j}, respectively, at the top of current page.
	\setlength{\textfloatsep}{0pt}
	\begin{table*}
		%	\vspace*{-0.6cm}
		\begin{normalsize} 
			\begin{equation}
			\label{F_j_1}
			\begin{split}
			\mathcal{F}_j^{(1)}\left(\mathbf{Z}, \mathbf{X}\right)=&\sum_{n \in \mathcal{N}_j}\sum_{i=1}^2 \omega_{j,l_j,n,}^{(1)} \left(\log\left(1+Z_{j,l_j,n}^{(i)}\right)-Z_{j,l_j,n}^{(i)}\right)
			+\sum_{n \in \mathcal{N}_j}\sum_{i=1}^3 \omega_{j,l_j,n,}^{(2)} \left(\log\left(1+X_{j,l_j,n}^{(i)}\right)-X_{j,l_j,n}^{(i)}\right),
			\end{split}
			\end{equation}
			\begin{equation}
			\label{F_j_2}
			\begin{split}
			&\mathcal{F}_j^{(2)}\left(\mathbf{P}, \mathbf{Q},\mathbf{Z}, \mathbf{X}\right)=\sum_{n \in \mathcal{N}_j}\sum_{i=1}^2 \frac{ \omega_{j,l_j,n,}^{(1)}\left(1+Z_{j,l_j,n}^{(i)}\right)A_{j,l_j,n}^{(i)}}{A_{j,l_j,n}^{(i)}+B_{j,l_j,n}^{(i)}}\\
			&+\sum_{n \in \mathcal{N}_j}\omega_{j,l_j,n}^{(1)}\log\left(1+\frac{ P_{j,l_j,n}^{(1,2)} \left|h_{j,l_j,n}^{(1)}\right|^2}{I_{c,l_j,n}}\right)
			+\sum_{n \in \mathcal{N}_j}\sum_{i=1}^3 \frac{ \omega_{j,l_j,n,}^{(1)}\left(1+X_{j,l_j,n}^{(i)}\right)\widehat{A}_{j,l_j,n}^{(i)}}{\widehat{A}_{j,l_j,n}^{(i)}+\widehat{B}_{j,l_j,n}^{(i)}},
			\end{split}
			\end{equation}
			\begin{equation}
			\label{Psi_j}
			\begin{split}
			\Psi_j\left(\mathbf{P},\mathbf{Q}\right)=&\sum_{n \in \mathcal{N}_j}\mu_n\left(\sum_{i=1}^3 \mathbf{P}_{j,l_j,n}[i]+\sum_{i=1}^3 \mathbf{Q}_{j,l_j,n}[i]\right)\\
			&	+\bm{\sigma_j}^T\left[\sum_{n \in\mathcal{N}_j}\sum_{i=1}^2  \mathbf{P}_{j,l_j,n}[i], \sum_{n \in\mathcal{N}_j} \mathbf{P}_{j,l_j,n}[3], \sum_{n \in\mathcal{N}_j}\sum_{i=1}^3  \mathbf{Q}_{j,l_j,n}[i] \right]^T.
			\end{split}
			\end{equation}
		\end{normalsize}
		%	\vspace*{-0.2cm}
		\hrulefill
	\end{table*}
	Moreover, $\omega_{j,l_j,n,}^{(1)}=\frac{1-\lambda_{j,l_j,n}}{2\log2}$, $\omega_{j,l_j,n,}^{(2)}=\frac{\lambda_{j,l_j,n}}{2\log2}$;
	$\lambda_{j,l_j,n} \in (0,1)$ is a parameter whose optimal value is obtained as such that $\mathrm{R}_{j,l_j,n}^{(U)}=\mathrm{R}_{j,l_j,n}^{(D)}$ is satisfied; and $I_{c,l_j,n}=\mathcal{P}_{c,n}|\tilde{h}_{c,l_j,n}|^2+\sigma^2$.  
\end{proposition}

\proof  The proof is provided in Appendix D.

By using primal-decomposition, we can decompose \eqref{BRS_II}  into outer and inner optimization problems. The outer and inner optimization problems are respectively given as  
\begin{equation}
\label{Outer}
\max_{\mathbf{Z} \geqq \bm{0}, \mathbf{X} \geqq \bm{0}} \mathcal{F}_j^{(1)}\left(\mathbf{Z}, \mathbf{X}\right) + \mathcal{F}_j^{(2)}\left(\mathbf{P}, \mathbf{Q},\mathbf{Z}, \mathbf{X}\right)
\end{equation}
and
\begin{equation}
\label{Inner}
\max_{\mathbf{P} \geqq \bm{0}, \mathbf{Q} \geqq \bm{0}} \mathcal{F}_j^{(2)}\left(\mathbf{P}, \mathbf{Q},\mathbf{Z}, \mathbf{X}\right)-\Psi_j \left(\mathbf{P},\mathbf{Q}\right).
\end{equation}
Moreover, by using the quadratic-transformation of fractional programming problem \cite[Theorem 1]{W_YU_II}, the inner optimization problem in \eqref{Inner} is further converted to a bi-convex optimization problem given as \eqref{Quadratic_Inner} at the top of current page. In \eqref{Quadratic_Inner}, $\bm{\alpha}=\left[\alpha_{j,l_j,n}^{(i)}\right]_{\substack{i \in \{1,2\} \\ n \in \mathcal{N}}}$ and  $\bm{\beta}=\left[\beta_{j,l_j,n}^{(i)}\right]_{\substack{i \in \{1,2,3\}\\ n \in \mathcal{N}}}$ are  the two sets of auxiliary variables introduced to transform the fractions to quadratic functions.
\setlength{\textfloatsep}{0pt}
\begin{table*}
	%\vspace*{-0.6cm}
	\begin{normalsize}
		\begin{equation}
		\label{Quadratic_Inner}
		\begin{split}
		\max_{\{\mathbf{P}, \mathbf{Q}, \bm{\alpha},\bm{\beta}\} \geqq 0}& \sum_{n \in \mathcal{N}_j}\sum_{i=1}^2 2\alpha_{j,l_j,n}^{(i)}\sqrt{\omega_{j,l_j,n}^{(1)}\left(1+Z_{j,l_j,n}^{(i)}\right)A_{j,l_j,n}^{(i)}} -\left(\alpha_{j,l_j,n}^{(i)}\right)^2\left(A_{j,l_j,n}^{(i)}+B_{j,l_j,n}^{(i)}\right)\\
		&+ \sum_{n \in \mathcal{N}_j}\sum_{i=1}^3 2\beta_{j,l_j,n}^{(i)}\sqrt{\omega_{j,l_j,n}^{(2)}\left(1+X_{j,l_j,n}^{(i)}\right)\widehat{A}_{j,l_j,n}^{(i)}} -\left(\beta_{j,l_j,n}^{(i)}\right)^2\left(\widehat{A}_{j,l_j,n}^{(i)}+\widehat{B}_{j,l_j,n}^{(i)}\right)\\
		&+\sum_{n \in \mathcal{N}_j} \omega_{j,l_j,n}^{(1)}\log\left(1+\frac{ P_{j,l_j,n}^{(1,2)} \left|h_{j,l_j,n}^{(1)}\right|^2}{I_{c,l_j,n}}\right)-\Psi_j\left(\mathtt{P},\mathtt{Q}\right).
		\end{split}
		\end{equation}
	\end{normalsize}
	\begin{subequations}
		\hrulefill
		\begin{align}
		&P_{j,l_j,n}^{(1,1)^*}=\frac{\omega_{j,l_j,n}^{(1)}\left(1+Z_{j,l_j,n}^{(1)}\right)\left(\alpha_{j,l_j,n}^{(1)}\left|h_{j,l_j,n}^{(1)}\right|\right)^2}{\left(\mu_n+\sigma_{j,1}+\left(\alpha_{j,l_j,n}^{(1)}\left|h_{j,l_j,n}^{(1)}\right|\right)^2\right)^2}, \forall n \in \mathcal{N}_j, l_j \in \mathcal{L} \label{CH_DU_power_1st_stream}\\
		&P_{j,l_j,n}^{(1,2)^*}= \left[\frac{\omega_{j,l_j,n}^{(1)}}{\mu_n+\sigma_{j,1}+\left|h_{j,l_j,n}^{(1)}\right|^2\sum_{i=1}^2\left(\alpha_{j,l_j,n}^{(i)}\right)^2}-\frac{I_{c,l_j,n}}{\left|h_{j,l_j,n}^{(1)}\right|^2}\right]^+, \forall n \in \mathcal{N}_j, l_j \in \mathcal{L} \label{CH_DU_power_2nd_stream}\\
		&P_{j,l_j,n}^{(2)^*}=\frac{\omega_{j,l_j,n}^{(1)}\left(1+Z_{j,l_j,n}^{(2)}\right)\left(\alpha_{j,l_j,n}^{(2)}\left|h_{j,l_j,n}^{(2)}\right|\right)^2}{\left(\mu_n+\sigma_{j,2}+\left|h_{j,l_j,n}^{(2)}\right|^2\sum_{i=1}^2\left(\alpha_{j,l_j,n}^{(i)}\right)^2\right)^2}, \forall n \in \mathcal{N}_j, l_j \in \mathcal{L}. \label{CH_DU_power_3rd_stream}
		\end{align}
	\end{subequations}
	\hrulefill
	%\vspace*{-0.6cm}
	%\hrulefill
\end{table*}

\begin{lem}
	\label{Corollary_Potenital_2}
	The NE  power allocation strategy of the $j$-th device-cluster is obtained by  solving \eqref{Outer} and \eqref{Quadratic_Inner} alternately, $\forall j \in \mathcal{T}$.
\end{lem}

\proof  The proof is provided in Appendix E.

The optimal solutions to \eqref{Outer} and \eqref{Quadratic_Inner} are summarized as follows.
\paragraph{Optimal $\{\mathbf{P^*}, \mathbf{Q^*}\}$}
For the given $\{\bm{\alpha,\beta}\}$ and $\{\mathbf{Z}, \mathbf{X}\}$, eq. \eqref{Quadratic_Inner} is a convex optimization problem of $\{\mathbf{P,Q}\}$. By satisfying the Karush-Khun-Tucker (KKT) optimality conditions to \eqref{Quadratic_Inner}, the CH DUs' optimal power allocations in the $j$-th device cluster, $\forall j \in \mathcal{T}$, are obtained as \eqref{CH_DU_power_1st_stream}-\eqref{CH_DU_power_3rd_stream} at the top of previous page   where $[a]^+=\max(a,0)$. Similarly, by satisfying the KKT optimality conditions, optimal power allocations of the $l_j$-th eRRH in the $j$-th device cluster, $\forall j \in \mathcal{T}$, are obtained as \eqref{relay_CU_power_1st_stream}-\eqref{relay_CU_power_3rd_stream} at the top of current  page. In \eqref{relay_CU_power_2nd_stream} and \eqref{relay_CU_power_3rd_stream}, $\mathrm{B}_{j,l_j,n}=\left(\beta_{j,l_j,n}^{(1)}g_{j,l_j,n}^{(\kappa)}\right)^2 + \left(\beta_{j,l_j,n}^{(2)}g_{j,l_j,n}^{(1)}\right)^2+ \left(\beta_{j,l_j,n}^{(3)}g_{j,l_j,n}^{(2)}\right)^2$. Here,  $\kappa \in \{1,2\}$ and it is obtained as $\kappa=\arg \min \left(\frac{|g_{j,l_j,n}^{(1)}|^2}{\mathcal{P}_{c,n}|\tilde{g}_{c,j,n}^{(1)}|^2+\sigma^2},\frac{|g_{j,l_j,n}^{(2)}|^2}{\mathcal{P}_{c,n}|\tilde{g}_{c,j,n}^{(2)}|^2+\sigma^2}\right)$. 
%is the index of CR DU with smaller channel-gain-to-interference-plus-noise ratio.
\setlength{\textfloatsep}{0pt}
\begin{table*}
	%	\vspace*{-0.6cm}
	\begin{normalsize}
		
		%\end{normalsize}
		%\vspace*{-0.4cm}
		%\hrulefill
		%\end{table*}
		%\begin{table*}
		%	\vspace*{-0.6cm}
		%	\begin{normalsize}
		\begin{subequations}
			\begin{align}
			&Q_{l_j,\bm{C},n}^*=\frac{\omega_{j,l_j,n}^{(2)}\left(1+X_{j,l_j,n}^{(1)}\right)\left(\beta_{j,l_j,n}^{(1)}\left|g_{j,l_j,n}^{(\kappa)}\right|\right)^2}{\left(\mu_n+\sigma_{j,3}+\left(\beta_{j,l_j,n}^{(1)}\left|g_{j,l_j,n}^{(\kappa)}\right|\right)^2\right)^2}, \forall n \in \mathcal{N}_j, l_j \in \mathcal{L} \label{relay_CU_power_1st_stream}\\
			& Q_{l_j,j,n}^{(1)^*}=\frac{\omega_{j,l_j,n}^{(2)}\left(1+X_{j,l_j,n}^{(2)}\right)\left(\beta_{j,l_j,n}^{(2)}\left|g_{j,l_j,n}^{(1)}\right|\right)^2}{\left(\mu_n+\sigma_{j,3}+\mathrm{B}_{j,l_j,n}\right)^2}, \forall n \in \mathcal{N}_j, l_j \in \mathcal{L}\label{relay_CU_power_2nd_stream}\\
			&	Q_{l_j,j,n}^{(2)^*}=\frac{\omega_{j,l_j,n}^{(2)}\left(1+X_{j,l_j,n}^{(3)}\right)\left(\beta_{j,l_j,n}^{(3)}\left|g_{j,l_j,n}^{(2)}\right|\right)^2}{\left(\mu_n+\sigma_{j,3}+\mathrm{B}_{j,l_j,n}\right)^2}, \forall n \in \mathcal{N}_j, l_j \in \mathcal{L}.\label{relay_CU_power_3rd_stream}
			\end{align}
		\end{subequations}
		\hrulefill
		\begin{subequations}
			\begin{align}
			&\alpha_{j,l_j,n}^{(i)^*}=\frac{\sqrt{\omega_{j,l_j,n}^{(1)}\left(1+Z_{j,l_j,n}^{(i)}\right)A_{j,l_j,n}^{(i)}}}{A_{j,l_j,n}^{(i)}+B_{j,l_j,n}^{(i)}}, \forall i=1,2, n \in \mathcal{N}_j, l_j \in \mathcal{L} \label{opt_alpha}\\
			&\beta_{j,l_j,n}^{(i)^*}=\frac{\sqrt{\omega_{j,l_j,n}^{(2)}\left(1+X_{j,l_j,n}^{(i)}\right)\widehat{A}_{j,l_j,n}^{(i)}}}{\widehat{A}_{j,l_j,n}^{(i)}+\widehat{B}_{j,l_j,n}^{(i)}}, \forall i=1,2,3, n \in \mathcal{N}_j, l_j \in \mathcal{L}. \label{opt_beta}
			\end{align}
		\end{subequations}
	\end{normalsize}
	%	\vspace*{-1.2cm}
	\hrulefill
\end{table*}

\paragraph{Optimal $\{\bm{\alpha^*}, \bm{\beta^*}\}$} For the given $\{\mathbf{P,Q}\}$ and $\{\mathbf{Z}, \mathbf{X}\}$,  eq. \eqref{Quadratic_Inner} is a convex optimization problem of $\{\bm{\alpha},\bm{\beta}\}$. By satisfying the KKT optimality conditions, the optimal $\{\bm{\alpha^*}, \bm{\beta^*}\}$, $\forall n\in \mathcal{N}_j, l_j \in \mathcal{L}, j \in \mathcal{T}$, are obtained as \eqref{opt_alpha} and \eqref{opt_beta} at the top of current page.

\paragraph{Optimal $\{\mathbf{Z}^*, \mathbf{X}^*\}$}  For the given $\{\mathbf{P,Q}\}$, eq. \eqref{Outer} is a strict convex optimization problem of  $\{\mathbf{Z}, \mathbf{X}\}$. By satisfying the KKT optimality conditions, $\forall n\in \mathcal{N}_j, l_j \in \mathcal{L}, j \in \mathcal{T}$, we obtain
\begin{equation}
\label{Opt_Z_X}
\begin{split}
Z_{j,l_j,n}^{(i)^*}=\frac{A_{j,l_j,n}^{(i)}}{B_{j,l_j,n}^{(i)}}, i=1,2 \quad \text{and} \quad 
X_{j,l_j,n}^{(i)^*}=\frac{\widehat{A}_{j,l_j,n}^{(i)}}{\widehat{B}_{j,l_j,n}^{(i)}}, i=1,2,3.
\end{split}
\end{equation}

\paragraph{Optimal $\{\bm{\lambda^*, \sigma^*}\}$}
A one-dimensional bi-section search in the $(0,1)$ interval is conducted to find the optimal $\{\lambda_{j,l_j,n}^*\}$ such that $\mathrm{R}_{j,l_j,n}^{(U)}=\mathrm{R}_{j,l_j,n}^{(D)}$ is satisfied,  $\forall n\in \mathcal{N}_j, l_j \in \mathcal{L}, j \in \mathcal{T}$.  \textcolor{black}{The pricing variables  $\{\sigma_{j,1},\sigma_{j,2}, \sigma_{j,3}\}$ in \eqref{Payoff} are  updated iteratively by applying the sub-gradient method as such  the constraint $\text{C4}$ is satisfied. For the brevity, the detailed analyses are omitted.}
\subsubsection{Power Allocation Algorithm}
We propose Algorithm \ref{Algorithm2} to obtain NE power allocation strategy for both hops in the $j$-th device-cluster assuming that the $j$-th device-cluster is assigned with the $l_j$-th eRRH and the $\mathcal{N}_j$ set of RRBs. The properties of Algorithm \ref{Algorithm2}  are discussed as follows.

\begin{proposition}
	\label{Proposition_Potential_Optimality}
	Algorithm \ref{Algorithm2} converges to a local-optimal solution  to $\text{P2.1}$.
\end{proposition}

\proof   The proof is provided in Appendix F.

\textit{Complexity:} To complete the Steps 5-9, Algorithm \ref{Algorithm2} requires in total $21|\mathcal{N}_j|$ computations. Since a sub-gradient method of updating the pricing variables requires $\mathcal{O}\left(\frac{1}{\epsilon^2}\right)$ iterations with $\epsilon$-accuracy, the overall computational complexity of Algorithm \ref{Algorithm2} is $\mathcal{O}\left(\frac{N}{\epsilon^2}\right)$.

\setlength{\textfloatsep}{0pt}
\begin{algorithm}
	\caption{NE Power Allocation Strategy  for the $j$-th Device-cluster}
	\label{Algorithm2}
	\begin{algorithmic}[1]
		\State \textbf{Input:} Assigned eRRH $l_j \in \mathcal{L}$ and set of RRBs $\mathcal{N}_j$; Maximum iterations $T_{max}$.
		
		\State \textbf{Initialize} $\{P_{j,l_j,n}^{(1,1)},P_{j,l_j,n}^{(1,2)}, P_{j,l_j,n}^{(2)}\}$ and $\{Q_{l_j, j,n}^{(1)},Q_{l_j,j,n}^{(2)}, Q_{l_j, \bm{C}, n}\}$, and calculate initial values of  $\{\alpha_{j,l_j,n}^{(i)}, \beta_{j,l_j,n}^{(i)}\}$ and $\{Z_{j,l_j,n}^{(i)}, X_{j,l_j,n}^{(i)}\}$ by using  \eqref{opt_alpha}, \eqref{opt_beta}, and \eqref{Opt_Z_X}, $\forall  n \in \mathcal{N}_j$.
		
		\State \textbf{Initialize} $\lambda_{j,l_j,n}^{low}=0$, $\lambda_{j,l_j,n}^{up}=1$, and $\lambda_{j,l_j,n}=\frac{\lambda_{j,l_j,n}^{low}+\lambda_{j,l_j,n}^{up}}{2}$; $\{\sigma_{j,1},\sigma_{j,2}, \sigma_{j,3}\}$; $t=1$;
		
		\Repeat
		
		\State Update  $\{P_{j,l_j,n}^{(1,1)},P_{j,l_j,n}^{(1,2)}, P_{j,l_j,n}^{(2)}\}$ and $\{Q_{l_j, j,n}^{(1)},Q_{l_j,j,n}^{(2)}, Q_{l_j, \bm{C}, n}\}$ by using \eqref{CH_DU_power_1st_stream}-\eqref{relay_CU_power_3rd_stream}, $\forall  n \in \mathcal{N}_j$.
		
		\State Update $\{Z_{j,l_j,n}^{(i)}, X_{j,l_j,n}^{(i)}\}$ by applying the updated power allocations to \eqref{Opt_Z_X}, $\forall  n \in \mathcal{N}_j$.
		
		\State Update  $\{\alpha_{j,l_j,n}^{(i)}, \beta_{j,l_j,n}^{(i)}\}$ by applying the updated power allocations  and the updated $\{Z_{j,l_j,n}^{(i)}, X_{j,l_j,n}^{(i)}\}$ to \eqref{opt_alpha} and \eqref{opt_beta}, respectively, $\forall  n \in \mathcal{N}_j$.
		
		\State If $\mathrm{R}_{j,l,n}^{(U)}>\mathrm{R}_{j,l,n}^{(D)}$, $\lambda_{j,l_j,n}^{up} \leftarrow \lambda_{j,l_j,n}$, and if $\mathrm{R}_{j,l,n}^{(U)}<\mathrm{R}_{j,l,n}^{(D)}$, $\lambda_{j,l_j,n}^{low} \leftarrow \lambda_{j,l_j,n}$. Update $\lambda_{j,l_j,n}=\frac{\lambda_{j,l_j,n}^{low}+\lambda_{j,l_j,n}^{up}}{2}$; $\forall  n \in \mathcal{N}_j$.
		
		\State Update $\{\sigma_{j,1},\sigma_{j,2}, \sigma_{j,3}\}$ by using the sub-gradient method; $t=t+1.$

		\Until{ $t>T_{max}$ or Convergence}
		
		\State \textbf{Output:} Power allocations for the first and second hops of the $j$-th device-cluster.
	\end{algorithmic}
\end{algorithm}

\textit{Implementation:} For a given device-cluster, the proposed framework requires power allocations for the  available eRRHs and RRBs. Thus, for each device-cluster a set of power allocations need to be  computed. To this end, both CH  DUs and CR DUs of a given device-cluster first estimate CSI of the available RRBs in the first and second hops, respectively. Next, both CH DUs and CR DUs send such estimated CSI to the available eRRHs. Finally, these eRRHs execute Algorithm \ref{Algorithm2}, and compute the required power allocations for the first and second hops of the considered device-cluster.

\vspace*{-0.4cm}
\subsection{Solution to $\text{P2.2}$: Resource Assignment by CBS}
\vspace{-0.2cm}
The power allocation variables need to be known to solve $\text{P2.2}$. For this purpose, by using the output of Algorithm \ref{Algorithm2}, each eRRH first computes the  achievable E2E rates and total transmit power over the available RRBs for the $j$-th device-cluster, $\forall j \in \mathcal{T}$. Next, each eRRH uploads such information to  CBS over  fronthaul links. Finally, using the received information from all the eRRHs, CBS constructs a  utility matrix for each device-cluster. Let $\mathbb{\widehat{U}}_j=\left[\widehat{U}_{l,n}^{(j)}\right] \in \mathbb{R}^{|\mathcal{L}| \times |\mathcal{N}|}$ be the utility matrix of the $j$-th device-cluster, where,
\begin{equation}
\label{U_j_l_n}
\widehat{U}_{l,n}^{(j)}=\left[R_{j,l,n}^{e2e}\left(\mathbf{P}_{j,l,n}^*,\mathbf{Q}_{j,l,n}^*\right)-\mu_n\left(\sum_{i=1}^3  \mathbf{P}_{j,l,n}^*[i]+\sum_{i=1}^3 \mathbf{Q}_{j,l,n}^*[i] \right)\right]^+.
\end{equation}
In \eqref{U_j_l_n}, $\{\mathbf{P}_{j,l,n}^*,\mathbf{Q}_{j,l,n}^*\}$ are the optimal power allocations of the $j$-th device-cluster over the $l$-th eRRH and the $n$-the RRB. Using the aforementioned utility matrices, we can express $\text{P2.2}$  as
\begin{equation}
\label{Optimization_Leader_1}
\begin{split}
& \max_{ \mathbf{x}\in \{0,1\}, \mathbf{y}\in \{0,1\} }\sum_{j \in \mathcal{T}} \sum_{l=1}^L\sum_{n=1}^N x_{l,j}y_{n,l,j}
\widehat{U}_{l,n}^{(j)} \quad \text{s.t.} \quad \text{ C2, C3}.
\end{split}
\end{equation}
Eq. \eqref{Optimization_Leader_1} is a bilinear integer programming problem, and it can be efficiently solved by using the alternating optimization \cite{Bilinear}. Given $\{y_{n,l,j}\}$, eq. \eqref{Optimization_Leader_1} can be simplified to
\begin{equation}
\label{Optimization_Leader_2}
\begin{split}
& \max_{ \mathbf{x}\in \{0,1\} }\sum_{j \in \mathcal{T}} \sum_{l=1}^L x_{l,j}
\widetilde{U}_{l}^{(j)} \quad \text{s.t.} \quad \text{ C2}
\end{split}
\end{equation}
where $\widetilde{U}_{l}^{(j)}=\sum_{n:y_{n,l,j}=1}\widehat{U}_{l,n}^{(j)}$. Eq. \eqref{Optimization_Leader_2} is an assignment problem and its optimal solution can be efficiently obtained in polynomial time by using the well-known Hungarian algorithm \cite{Hungarian}. Applying the solution to \eqref{Optimization_Leader_2} to \eqref{Optimization_Leader_1}, we can express  \eqref{Optimization_Leader_1} to
\begin{equation}
\label{Optimization_Leader_3}
\begin{split}
& \max_{ \mathbf{\tilde{y}}\in \{0,1\} }\sum_{j \in \mathcal{T}} \sum_{n=1}^N \tilde{y}_{n,j}
\breve{U}_{n}^{(j)} \quad \text{s.t.} \quad \text{ C3}
\end{split}
\end{equation}
where $\breve{U}_{n}^{(j)}=\sum_{l: x_{l,j}=1}\widehat{U}_{l,n}^{(j)}$ and $\tilde{y}_{n,j}$ is a new binary variable, i.e.,  $\tilde{y}_{n,j}=1$ if the $n$-th RRB is assigned to the $j$-th device-cluster, and $\tilde{y}_{n,j}=0$ otherwise. 
For a given solution to \eqref{Optimization_Leader_2}, the optimal solution to \eqref{Optimization_Leader_3} is obtained as
\begin{equation}
\label{Solution_2}
\tilde{y}_{n,j}^*=
\begin{cases}
1, & \mbox{if}  \quad j=\arg\max_{\substack{ j' \in \mathcal{T}}} \breve{U}_{n}^{(j')} \\
0, & \mbox{otherwise}.
\end{cases}
\end{equation}
Subsequently, we recover $\{y_{n,l,j}\}$ as  $y_{n,l,j}=x_{l,j}\tilde{y}_{n,j}^*$, $\forall n,l,j$.
By solving \eqref{Optimization_Leader_2} and  \eqref{Optimization_Leader_3} iteratively,  we obtain a convergent solution to $\eqref{Optimization_Leader_1}$. The overall steps are summarized in Algorithm \ref{Algorithm3}. The local-optimality and computational complexity of Algorithm \ref{Algorithm3}  are discussed as follows.

\vspace{-0.5cm}
\begin{algorithm}
	\caption{Assignment of RRBs and eRRHs among the Device-clusters}
	\label{Algorithm3}
	\begin{algorithmic}[1]
		\State \textbf{Input:} Utility matrix, $\mathbb{\widehat{U}}_j$, $\forall j \in \mathcal{T}$; Maximum iterations $T_{max}$.
		
		\State \textbf{Initialize:} $\{y_{n,l,j}\}, \forall n,l,j$; $t=1$.
		
		\Repeat
		\State Calculate  $\widetilde{U}_{l}^{(j)}$, $\forall l,j$ by using the  current value of $\{y_{n,l,j}\}$.
		
		\State Solve \eqref{Optimization_Leader_2} by using the Hungarian algorithm  of \cite{Hungarian} and update $\{x_{l,j}\}$.
		
		\State Calculate $\breve{U}_{n}^{(j)}$ by using the updated $\{x_{l,j}\}$.
		
		\State Calculate $\tilde{y}_{n,j}^*$ by using \eqref{Solution_2} and update $\{y_{n,l,j}\}$; $t=t+1$
		
		\Until{$\sum_{j \in \mathcal{T}} \sum_{l=1}^L\sum_{n=1}^N x_{l,j}y_{n,l,j}
			\widehat{U}_{l,n}^{(j)} $ converges or $t> T_{max}$}	
		
		\State \textbf{Output:} $x_{l,j}^*$ and  $y_{n,l,j}^*$, $\forall n,l,j$.
	\end{algorithmic}
\end{algorithm}
\vspace{-0.4cm}

\begin{proposition}
	\label{Prop_Leader_1}
	Algorithm \ref{Algorithm3} provides a local-optimal solution to \eqref{Optimization_Leader_1}.
\end{proposition}

\proof   The proof is provided in Appendix G.

\textit{Computational complexity:} Assuming that the number of device-clusters is more than the number of eRRHs,  $\mathcal{O}(|\mathcal{T}|^3)$ iterations are required to solve \eqref{Optimization_Leader_2} in Step 5 of Algorithm \ref{Algorithm3} \cite{Hungarian}. Meanwhile, a  total of $N|\mathcal{T}|$  operations are required  to execute Step 7 in Algorithm \ref{Algorithm3}. Since $|\mathcal{T}|=\lceil\frac{M}{2}\rceil$, the overall computational complexity of Algorithm  \ref{Algorithm3} is  $\mathcal{O}(|\mathcal{T}|^3+ |\mathcal{T}|N))=\mathcal{O}\left(\lceil\frac{M}{2}\rceil^3 +\lceil\frac{M}{2}\rceil N\right)$.

\vspace*{-0.3cm}
\subsection{Solution to $\text{P2.3}$: Pricing of the RRBs by CBS}
%\vspace*{-0.2cm}
To solve $\text{P2.3}$  optimally, CBS requires the global CSI and the optimization parameters of the entire network. However, in our considered system, CBS only knows the  utility matrices of  the device-clusters, uploaded by the eRRHs,  and the assignment of the eRRHs and RRBs among the device-clusters. Hence,  a solution to $\text{P2.3}$ can be obtained from the following feasibility problem
\begin{equation}
\label{price}
\begin{split}
& \text{find} \quad \mu_n, \forall n \in \mathcal{N}_{sc} \quad \text{s.t.} \quad \sum_{j \in \mathcal{T}}\sum_{l=1}^L x_{l,j}^*y_{n,l,j}^* \max\left(I_{j,l,n}^{(1)}, I_{j,l,n}^{(2)}\right) \leq I_{th},  \forall n \in \mathcal{N}_{sc}
\end{split}
\end{equation}
where $\{x_{l,j}^*, y_{n,l,j}^*\}$ are obtained from the output of  Algorithm \ref{Algorithm3}. The uplink interference at the $n$-th RRB becomes minimum (i.e., zero) and maximum for $\mu_n \to \infty$ and $\mu_n \to 0$, respectively. Hence, certain $\mu_n$ must exist for which  $I_n^{(up)}=I_{th}$ is satisfied where $I_n^{(up)}=\sum_{j \in \mathcal{T}}\sum_{l=1}^L x_{l,j}^*y_{n,l,j}^* \max\left(I_{j,l,n}^{(1)}, I_{j,l,n}^{(2)}\right)$. The solution to \eqref{price} is obtained by using a  bi-section search.  Let the resource assignments among the device-clusters be given, and let  $\mu_n^{(low)}$ and $\mu_n^{(up)}$ be the lowest and highest price of the $n$-th RRB, respectively. The initial prices of the RRBs are set as $\mu_n=\frac{\mu_n^{(low)}+\mu_n^{(up)}}{2}$, $\forall n \in \mathcal{N}_{sc}$. By applying the given prices of the RRBs to Algorithm \ref{Algorithm2}, the power allocations of the transmitting nodes are updated, and subsequently, the value of  $I_n^{(up)}$, $\forall n \in \mathcal{N}_{sc}$, is updated. If $I_n^{(up)}>I_{th}$,  $\mu_n^{(low)} \leftarrow \mu_n$ is applied, and $I_n^{(up)}<I_{th}$,  $\mu_n^{(up)} \leftarrow \mu_n$ is applied. Then, the  price of the $n$-th RRB is updated as $\mu_n=\frac{\mu_n^{(low)}+\mu_n^{(up)}}{2}$, $\forall n \in \mathcal{N}_{sc}$. The aforementioned procedure is repeated until $|I_n^{(up)}-I_{th}|$ approaches a  small value, $\forall n \in \mathcal{N}_{sc}$.

\section{Development and Properties of RSMD Algorithm}
\subsection{Overview of RSMD Algorithm}
The overall steps of the proposed RSMD algorithm  to obtain a suitable solution to $\text{P0}$ is summarized in Algorithm \ref{RSMD}. \textcolor{black}{RSMD is a multi-stage algorithm.} In  Stage-I, by using Algorithm \ref{Algorithm1}, the device-clusters are established. In  Stage-II, by using the initial prices of RRBs in Algorithm \ref{Algorithm2}, the power allocations for each device-cluster, eRRH and RRB are calculated, and by plugging such power allocations to Algorithm \ref{Algorithm3},  both eRRHs and RRBs are assigned for the device-clusters.  In  Stage-III,  based on the assigned resources, the power allocations of all the CH DUs and active eRRHs
are refined so that the sum-rate of each device-cluster is maximized and the uplink interference constraint at the CBS is satisfied. The resource allocations in  Stage-III, i.e.,  Steps 12-14 of Algorithm \ref{RSMD},
are  repeated iteratively until the maximum number of iterations is reached or the interference constraint is satisfied.  

Note that  the  aforementioned three stages  of  RSMD  require  semi-distributed implementation. In Stage-I,  the device-clusters are formed by utilizing both the device and cloud level computation as mentioned in Section III. B(1).  In Stage-II, the resource assignments among the device-clusters are obtained by exchanging information between the eRRHs and CBS over the fronthaul links as mentioned in Section IV. B. Finally, in Stage-III,  the power allocations are  updated distributively at the eRRHs, i.e., each eRRH only updates the power allocations for the assigned device-clusters; CBS updates prices of all the RRBs; and CBS shares the updated prices with the eRRHs over the fronthaul links.  
	
Moreover,   to perform the resource allocation among the device-clusters, RSMD  does not require global CSI at CBS and excessive signaling overhead.  Particularly, for executing Algorithm \ref{Algorithm3} in   Stage-II, maximum $\mathcal{O}\left(\lceil\frac{M}{2}\rceil LN\right)$ information exchanges are required between the eRRHs and CBS over the fronthaul links.  At each iteration of  Stage-III, maximum $\mathcal{O}\left(N\right)$ information exchanges are required  for sharing the updated RRB prices with the eRRHs over the fronthaul links.  Assuming that the total number of iterations of the Stage-III is $I_{max}$,  the RSMD algorithm requires in total $\mathcal{O}\left(\lceil\frac{M}{2}\rceil LN+I_{max} N\right)$ information exchanges over the fronthaul links. We emphasize that the required information exchanges  for implementing the proposed RSMD algorithm  is comparable with the standard resource allocation algorithms for F-RAN and C-RAN \cite{D2D_C-RAN}.  
%The implementation complexity of  the  RSMD algorithm can be further reduced by updating  Stage-I and Stage-II resource allocations based on the long-term CSI, and updating only Stage-III resource allocation based on the instantaneous CSI. 
Consequently,  the proposed RSMD algorithm is suitable for the practical implementation.

\vspace*{-0.3cm}
\subsection{Convergence of the Resource Allocation}
\textcolor{black}{At  Stage-III of  the proposed RSMD algorithm, the power allocations and  the RRBs' prices are  updated iteratively by using the notion of a Stackelberg game. Recall, the target of  a Stackelberg game-based resource allocation scheme  is to achieve the SE.  Particularly, let $\{\bm{P}^*, \bm{Q}^*\}$ be the final power allocations of the device-clusters, and $\bm{\mu^*}$ be the final prices of the RRBs.  Such a resource allocation will be the  SE if both the device-clusters and CBS have no incentive to change the power allocations and  the RRBs' prices from  $\{\bm{P}^*, \bm{Q}^*\}$ and $\bm{\mu^*}$, respectively \cite{D2D_C-RAN}.}

\begin{proposition}
	RSMD algorithm converges to the SE resource allocation strategy.
\end{proposition}

\proof  The proof is provided in Appendix H.

\textcolor{blue}{\textit{Remark 3}: The final resource allocation obtained by  the RSMD algorithm may not be a global optimal solution to  $\text{P2}$. The reasons are two-folds.  First,  the resource allocation at Stage-II of RSMD, i.e., the assignment of eRRHs and RRBs among the device-cluster is obtained by solving a bi-linear optimization problem iteratively. Therefore, despite the convergence, Stage-II resource allocation may only achieve a local optimal solution. Moreover, the overall optimization variables in $\text{P2}$  are decoupled between Stage-II and Stage-III of the RSMD algorithm. Such a decoupling also affects the optimality of  the resultant solution. Accordingly, the final resource allocation output of the RSMD algorithm is only  near-optimal. However, since $\text{P2}$ is  NP-hard, a converged and near-optimal solution to $\text{P2}$ is acceptable in practice.}
\vspace*{-0.2cm}
\begin{algorithm}
	\caption{Rate-splitting for Multi-hop D2D (RSMD) Algorithm}
	\label{RSMD}
	\begin{algorithmic}[1]
		\State \textbf{Input:} Set of  eigen-vectors for  CH DU and CR DU of each D2D link. (\textcolor{blue}{Start of  Stage-I}) 
		
		\State CH DU and CR DU of each D2D link compute PCVs  and upload these PCVs to CBS.
		
		\State CBS forms device-clusters by using Algorithm \ref{Algorithm1}.
		
		\State \textbf{Output:} Set of device-clusters. (\textcolor{blue}{End of  Stage-I}) 
		
		\State \textbf{Initialize:} CBS announces initial RRB prices. (\textcolor{blue}{Start of Stage-II}) 
		
		\State Using Algorithm \ref{Algorithm2}, an eRRH computes the power allocations for all the device-clusters and RRBs. Repeat for all the eRRHs.
		
		\State Each eRRH computes a set of utility metrics for the device-clusters based on the available power allocations, and uploads such metrics to the CBS over fronthaul.
		
		\State Based on the received utility metrics, CBS executes Algorithm \ref{Algorithm3} to determine assignment of the eRRHs and RRBs for the device-clusters.
		
		\State \textbf{Output:} Assignment of the eRRH and RRBs for the device-clusters. (\textcolor{blue}{End of  Stage-II}) 
		
		\State \textbf{Initialize:} Iteration index $i=1$, total number of iterations, $I_{max}$. (\textcolor{blue}{Start of Stage-III}) 
		
		\Repeat
		\State By plugging the updated RRB prices to Algorithm \ref{Algorithm2}, each eRRH computes the power allocations for the associated device-clusters and RRBs.

		\State CBS measures uplink interference at the RRBs and updates the prices of all the RRBs  by using the bi-section search of Section IV. C.
		
		\State CBS broadcasts the updated RRB prices over the fronthaul links; $i=i+1$.
		
		\Until{$i >I_{max}$ or the interference constraint at CBS is satisfied}
		
		\State \textbf{Output:} Power allocations for the device-clusters. (\textcolor{blue}{End of Stage-III})

	\end{algorithmic}
\end{algorithm}
%\vspace*{-1.2cm}

%\vspace*{-0.1cm}
\subsection{Computational Complexity}
Considering $\epsilon$ is the error tolerance level, the bi-section method of updating RRBs' prices require $\mathcal{O}\left(\frac{N}{\epsilon}\right)$ iterations. Hence, based on  the reported complexity of Algorithms \ref{Algorithm2} and \ref{Algorithm3}, the worst-case computational complexity for executing  Stage-II and  Stage-III of the RSMD  algorithm is  obtained as $\mathcal{O}\left(\frac{N}{\epsilon}\left(\lceil\frac{M}{2}\rceil\frac{NL}{\epsilon^2}\right)+\lceil\frac{M}{2}\rceil N+\lceil\frac{M}{2}\rceil^3 \right)$, and it is approximated as $\mathcal{O}\left(\lceil\frac{M}{2}\rceil\frac{N^2L}{\epsilon^3}+\lceil\frac{M}{2}\rceil^3 \right)$. 
Combining the complexity of both device-clustering and resource allocation phases, the overall computational complexity of the  RSMD algorithm is obtained as  $\mathcal{O}\left(\lceil\frac{M}{2}\rceil\frac{N^2L}{\epsilon^3}+\lceil\frac{M}{2}\rceil^3+MN^3 \right)$. Obviously, our proposed RSMD algorithm has polynomial computational complexity. 

\vspace{-0.3cm}
\section{Numerical Simulation Results}
For simulations, we consider an F-RAN with one CBS,  $50$ D2D links, $10$ eRRHs, and $36$ RRBs. Moreover, $36$  uplink CUEs are considered  assuming that each CUE occupies one  RRB and the maximum transmit power of each CUE is $0.5$ Watt. \textcolor{black}{The distances of  all the links in the system are  generated randomly by using the Poisson distribution. Particularly, the mean distance between  the eRRH and  the CH DU (and the CR DU) is set as $100m$;  the mean distance between  the eRRH and the CBS is set as $200m$;  the mean distance between  the CH DU and the CBS is set as $200m$;  the mean distance between the eRRH and the CUE is set as $160m$; and  the mean distance between the CUE and the CR DU is set as $230m$.}
Without further specification, we  consider $P_{max}^{(D)}=P_{max}^{(R)}=0.5$ Watt, $I_{th}=-80$ dBm,  and $\sigma^2=-174$ dBm. \textcolor{black}{In addition, to implement Algorithm \ref{Algorithm1} for all the simulations (except Fig. 6a), we consider $d=4$ PCVs and $w=0.5$.} Finally, Rayleigh distributed fading and 3GPP path loss model, given by, $131.1+42.8\log_{10}(d)$ dB (where $d$ is the distance in kilometers) is used to generate the channel coefficients for all the links in the system.
For  performance comparison, we consider the following benchmark schemes.

\begin{itemize}
	\item \textbf{PD-NOMA:} \textcolor{black}{ In this scheme, in stead of RS,  the conventional PD-NOMA is applied to mitigate the intra-device cluster interference. To this end, uplink and downlink PD-NOMA protocols are applied to obtain the transmit power allocations for the first and second hops of each device-cluster, respectively.  For a fair comparison, the device-clusters and the  resource assignments among the device-clusters are obtained by using  Algorithm \ref{Algorithm1} and Algorithm \ref{Algorithm3}, respectively. To obtain the transmit power allocations by using PD-NOMA, we consider three different techniques, namely, fractional programming \cite{W_YU_II}, successive convex approximation (SCA) \cite{Matching_1}, and iterative water-filling (IWF) \cite{Ruksansa_NOMA}, and  in the ensuing discussions, these PD-NOMA schemes are referred as  ``PD-NOMA/OPT'', ``PD-NOMA/SCA'', and ``PD-NOMA/IWF'', respectively.  }
	
	\item \textbf{Treat interference as noise (TIN) and multicasting (TIN/Multicasting):}  \textcolor{blue}{In this scheme, instead of RS, the TIN and multicasting protocols are used  at the first and second hops of each device-cluster, respectively. In the first hop of each device-cluster,   the eRRH decodes the transmitted message of both CH DUs by treating interference completely as noise, i.e., the eRRH does not perform any successive decoding and interference cancellation. In the second hop of each device-cluster, the eRRH transmits message to both CR DUs by employing a multicasting protocol, and thus, entirely eliminates interference between the CR DUs in the second hop. To maximize the end-to-end  sum-rate, the fractional programming based power control is applied at both hops.  For a fair comparison, the device-clusters and the  resource assignment among the device-clusters are obtained by using  Algorithm 1 and Algorithm  3, respectively.}

	\item  \textbf{Fixed resource allocation with water-filling power allocation (FRA/WF-PA):}  \textcolor{black}{This scheme entirely eliminates both intra device-cluster and inter device-cluster interference by scheduling each D2D link over the orthogonal RRBs. The RRBs are equally allocated among the active D2D links. The associations among  eRRHs and D2D links are obtained by applying a many-to-one matching algorithm.  Finally, to efficiently utilize the allocated RRBs,  water-filling power allocation (WF-PA)  is applied to maximize data rate of both hops of a D2D link.}
\end{itemize}

\begin{figure}
	\vspace*{-0.4cm}
	\begin{center}
		\subfloat[t][\textcolor{black}{E2E sum-rate  versus (vs.) number of  the D2D links}]{
			\includegraphics[width=0.45\linewidth, draft=false]{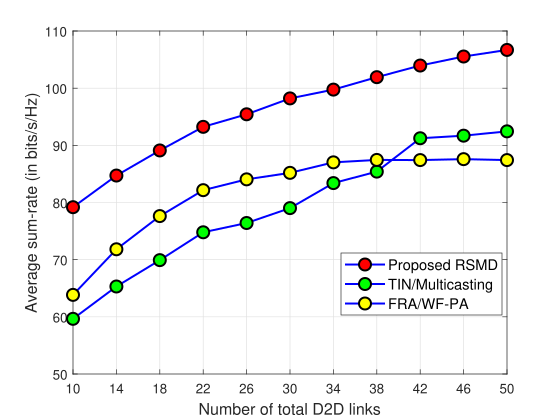}
			\label{fig_2_a}}
		\quad
		\subfloat[t][\textcolor{black}{E2E sum-rate vs. number of the D2D links}]{
			\includegraphics[width=0.45\linewidth, draft=false]{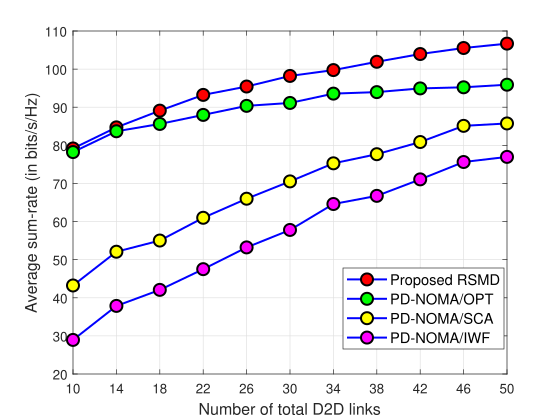}
			\label{fig_2_b}}
		
		\caption{\textcolor{black}{E2E sum-rate comparison between the proposed RSMD and benchmark schemes  considering $L=5$ eRRHs, $N=36$ RRBs, and $N_R=5$ device-clusters per eRRH} }
	\end{center}
	\vspace*{-0.2cm}
\end{figure}      

\vspace*{-0.5cm}
\subsection{Performance comparison for different D2D links}
\textcolor{black}{Fig. \ref{fig_2_a} compares  the average sum-rate of the proposed RSMD, TIN/Mutlicasting, and FRA/WF-PA schemes for different D2D links. Note that  FRA/WF-PA outperforms TIN/Multicasting for small number of D2D links. Such an observation is intuitively expected. As the number of D2D links is smaller than the number of RRBs, it is efficient to schedule each D2D link over the orthogonal RRB(s). Therefore,  device-clustering is not beneficial unless the intra device-cluster interference is properly managed. The \textit{interference-free} approach of FRA/WF-PA can support maximum $N$ number of the D2D links where  $N$ is the number of the available orthogonal RRBs in the system. Intuitively, TIN/Multicasting outperforms FRA/WF-PA as the number of D2D links exceeds the number of RRBs, thanks to the device-clustering. \textcolor{blue}{However, TIN/Multicasting does not 
remove any interference by applying an SIC technique at the eRRH during decoding the  messages received from  CH DUs. Hence, in the first hop of a given device-cluster, an uplink RS scheme achieves superior sum-rate than the TIN protocol \cite{RS_Uplink}. Moreover, TIN/Multicasting transmits data from the eRRH to both CR DUs by using a multicasting technique. Note that the achievable data rate in multicasting is dominated by the minimum rate of  the CR DUs. As a result,  in the second hop of each device-cluster,  a downlink RS scheme achieves  superior sum-rate than a multicasting protocol. Consequently, RSMD outperforms TIN/Multicasting in terms of the average sum-rate.} Meanwhile, even for small D2D links,  RSMD is benefited from device-clustering by efficiently managing both intra device-cluster and inter-device cluster interference. As a result, for all the D2D links in the system, RSMD achieves large average sum-rate gain over both benchmark schemes. Fig. \ref{fig_2_a} depicts that for $50$ D2D links, RSMD achieves $15.4\%$ and $22.04\%$ average sum-rate gain over TIN/Multicasting and FRA/WF-PA, respectively. Overall, RSMD is efficient than both TIN/Multicasting and FRA/WF-PA for the considered system.}

%RSMD is efficient since it allows  increased interference in the system and exploits such interference   through the optimized resource allocation.}

\textcolor{blue}{Fig. \ref{fig_2_b} compares the average sum-rate between the proposed RSMD and different PD-NOMA schemes by varying the number of D2D links. Fig. 2b depicts that RSMD outperforms  the PD-NOMA/OPT scheme for all the D2D links. The reason is explained as follows. In each device cluster, the first and second hops are the  two-user SISO MAC channel and SISO BC channel, respectively.  As per \cite[Lemma 6]{RS_CMD_1}, both RS and PD-NOMA enabled schemes achieve the optimal sum-rate of a two-user SISO BC channel. In contrast, unlike PD-NOMA, RS achieves the entire optimal rate region of  a two-user SISO MAC channel \cite{Walid_Sadd_Uplink_RSMA}. Recall, the E2E sum-rate of a device-cluster is determined by the minimum transmission rate of the first and second hops. As a result, for a given device-cluster, RS based transmission strategy always achieves same or better E2E sum-rate than PD-NOMA based transmission strategy. Therefore, RSMD provides average sum-rate gain over  PD-NOMA/OPT. Moreover, since RSMD can achieve the optimal rate region for each device-cluster, the average sum-rate gap between RSMD and PD-NOMA/OPT is enhanced substantially for a large number of D2D links (i.e., device-clusters), thanks to the  multi-user diversity. Fig. 2b depicts that RSMD achieves $1.21\%$ and $11.24\%$ average sum-rate gain over PD-NOMA/OPT for 10 and 50 D2D links, respectively. Recall, RSMD employs a fractional programming based transmit power allocation, which maximizes the sum-rate over the interference channels \cite{W_YU_II}. In contrast, both SCA and IWF based transmit power allocations are sub-optimal, and they may converge to an inferior solution. Consequently,  because of using the sub-optimal transmit power allocation, both PD-NOMA/SCA and PD-NOMA/IWF  achieve a considerably reduced average sum-rate than RSMD.  Fig. 2b depicts that for $50$ D2D links,  RSMD outperforms PD-NOMA/SCA and PD-NOMA/IWF by $24.43\%$ and $38.61\%$, respectively. Finally, note that the device-clusters are allocated with the orthogonal RRBs, and the number of total RRBs is finite. Moreover, the total transmit power of both CH DUs and eRRHs is also constrained by the maximum transmit power limit. As a result, there is a finite maximum average sum-rate of the system. The fractional programming based transmit power allocation, leveraging the near-optimality over an interference channel, can achieve a  high percentage of such a maximum average sum-rate even with a small number of the D2D links.  Hence, Fig. \ref{fig_2_b}  also depicts that RSMD has much higher average sum-rate gain over both PD-NOMA/SCA and PD-NOMA/IWF in small number of D2D links.  We conclude that for the considered system,  RSMD  is more efficient than  the conventional PD-NOMA.}

\vspace*{-0.3cm}
\subsection{Performance comparison for different eRRHs and RRBs}

\begin{figure}
	\vspace*{-0.4cm}
	\begin{center}
		\includegraphics[width=0.45\linewidth, draft=false]{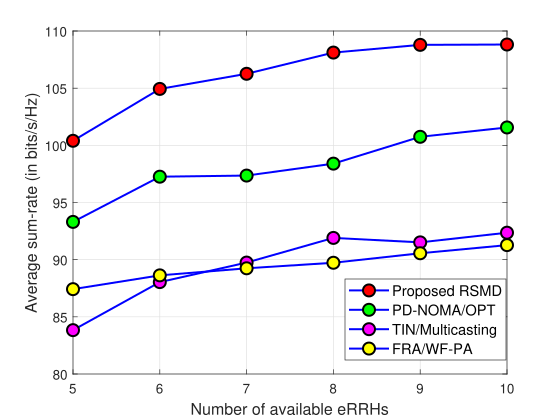}
		\caption{\textcolor{black}{E2E sum-rate comparison between the proposed  and benchmark schemes for different eRRHs  considering $M=36$ D2D links, $N=36$ RRBs, and $N_R=5$ device-clusters per eRRH} }
		\label{fig_3}
	\end{center}
	\vspace*{-0.2cm}
\end{figure}     
\begin{figure}
	\vspace*{-0.4cm}
	\begin{center}
		\subfloat[t][\textcolor{black}{E2E sum-rate  vs. the number of RRBs}]{
			\includegraphics[width=0.45\linewidth, draft=false]{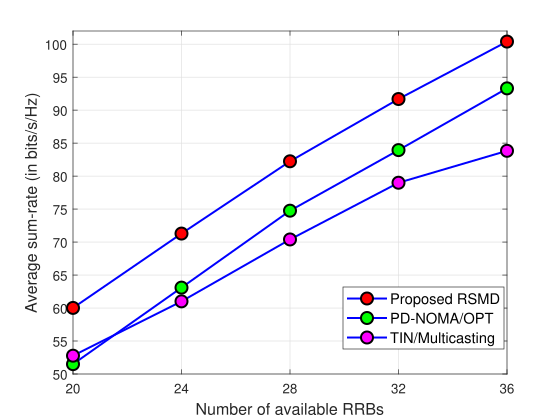}
			\label{fig_4_a}}
		\quad
		\subfloat[t][\textcolor{black}{E2E sum-rate vs. the number of RRBs}]{
			\includegraphics[width=0.45\linewidth, draft=false]{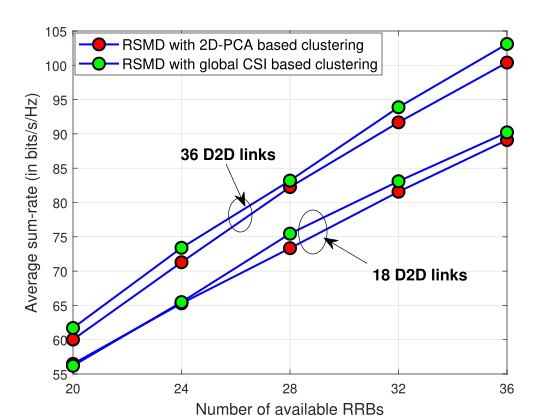}
			\label{fig_4_b}}
		\caption{\textcolor{black}{E2E sum-rate comparison between the proposed RSMD and benchmark schemes  considering $M=36$ D2D links, $L=5$ eRRHs, and $N_R=5$ device-clusters per eRRH} }
	\end{center}
	\vspace*{-0.2cm}
\end{figure} 

\textcolor{black}{Fig. \ref{fig_3} compares the average sum-rate of the proposed RSMD and benchmark schemes for different number of eRRHs in the system. As the number of eRRHs is increased, the probability of selecting the most suitable eRRH for the device-clusters (or D2D links) is enhanced. Obviously, the average sum-rate of all the schemes is improved  as the number of eRRHs is increased. Fig.  \ref{fig_3} also depicts that  FRA/WF-PA achieves almost similar average sum-rate to TIN/Multicasting. Such an observation supports the previously demonstrated fact, that is, as the number of D2D links is  same or small compared to the available RRBs, device-clustering is not always beneficial. In fact,  the intra device-cluster interference may  affect the sum-rate adversely,  and  efficient resource allocation is imperative to reap the benefit of device-clustering. Our proposed RSMD scheme adopts device-clustering and uses  efficient resource allocation to suppress the resultant interference. Consequently, RSMD achieves appreciable average sum-rate gain over the benchmark schemes for both small and large eRRHs in the system as depicted from Fig. \ref{fig_3}. For example,   for $L$=8 eRRHs in the system, RSMD achieves  $9.14\%$, $18.39\%$ and $19.06\%$ average sum-rate  gain over the PD-NOMA/OPT, TIN/Multicasting, and FRA/WF-PA schemes, respectively.}

\textcolor{black}{Fig. \ref{fig_4_a} plots the average sum-rate of the RSMD, PD-NOMA/OPT, and TIN/Multicasting schemes while varying the number of RRBs. RSMD  outperforms PD-NOMA/OPT  for all the RRBs, and the reason is explained in  the discussion   of  Fig. \ref{fig_2_b}.  Meanwhile, in both hops of a given device-cluster, TIN/Multicasting mitigates the intra device-cluster interference by using a sub-optimal approach.  Hence, both RSMD and PD-NOMA/OPT  outperform TIN/Multicasting, especially for large number RRBs. Fig. \ref{fig_4_a} depicts that for $N=32$ RRBs,   RSMD achieves $9.22\%$ and $16.07\%$  average sum-rate gain over  PD-NOMA/OPT and  TIN/Multicasting, respectively.}

\textcolor{black}{Fig. \ref{fig_4_b} plots the average sum-rate of RSMD for $18$ and $36$ D2D links while varying the number of RRBs. Both  the proposed 2D-PCA based device-clustering and the global CSI based device-clustering methods are considered. In the global CSI based device-clustering,  CBS is assumed to have the entire channel matrices for all the D2D links to form the device-clusters. Intuitively, the global CSI based device-clustering can achieve better sum-rate than proposed 2D-PCA based device-clustering method. However, Fig. \ref{fig_4_b} depicts that the performance gap between these two device-clustering methods is small. For instance, for $N=36$ RRBs, the proposed 2D-PCA based device-clustering achieves $98.75\%$ and $97.37\%$ of the average sum-rate of  the global CSI based device-clustering method, respectively. Recall, in the considered example, only four PCVs are used for performing the 2D-PCA based device-clustering. Obviously, for $N=36$ RRBs,  2D-PCA based device-clustering method requires $88.89\%$ less training overhead compared to a global CSI based device-clustering method. Therefore, the proposed 2D-PCA based device-clustering method  can  reduce the training overhead without  affecting the sum-rate.}

\begin{figure}
	\vspace*{-0.4cm}
	\begin{center}
		\subfloat[t][\textcolor{black}{E2E sum-rate vs. interference thresholds at CBS}]{
			\includegraphics[width=0.45\linewidth, draft=false]{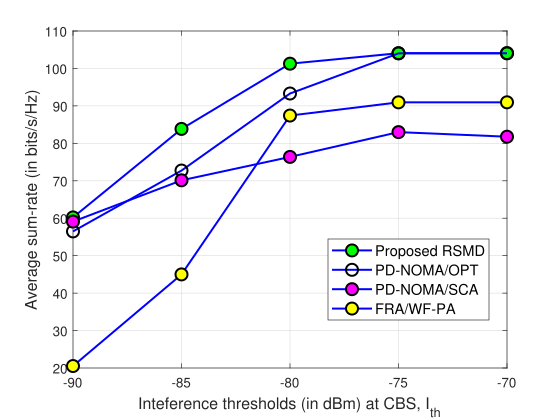}
			\label{fig_5_a}}
		\quad
		\subfloat[t][\textcolor{black}{Sum-transmit power vs. interference thresholds at CBS}]{
			\includegraphics[width=0.45\linewidth, draft=false]{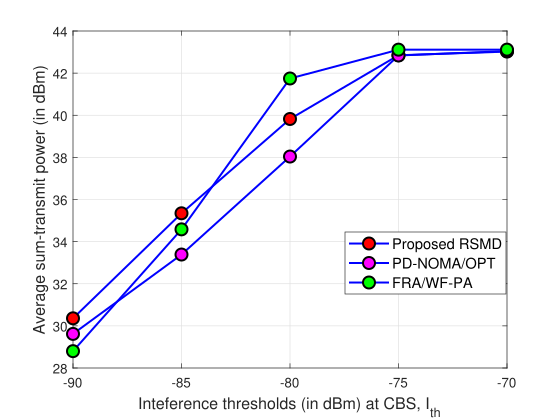}
			\label{fig_5_b}}
		
		\caption{\textcolor{black}{Rate and transmit power comparison between the proposed RSMD and benchmark schemes  for $M=36$ D2D links, $L=5$ eRRHs, $N=36$ RRBs,  and $N_R=5$ device-clusters per eRRH} }
	\end{center}
	\vspace*{-0.2cm}
\end{figure}

\vspace*{-0.3cm}
\subsection{Performance comparison for different  interference thresholds at CBS}
\textcolor{black}{Fig. \ref{fig_5_a} compares the average sum-rate of the proposed and benchmark schemes for different interference thresholds at CBS. As the interference threshold is increased, the CBS can tolerate more interference in uplink, and consequently, both the transmit power and sum-rate of the D2D network are increased.  Specifically, for a large interference threshold at the CBS,  all the CH DUs and active eRRHs can operate with their maximum transmit power, and thus, further increase of interference threshold does not  alter the power allocation of the transmitting nodes significantly. 
	For this reason,  the average sum-rate of  all the schemes in Fig. \ref{fig_5_a} is saturated in the large interference threshold regime. Fig. \ref{fig_5_a}  depicts that RSMD achieves  appreciable sum-rate gain over all the benchmark schemes, especially for moderate interference thresholds at the CBS. For instance, at $I_{th}=-80$ dBm,  RSMD achieves $8.54\%$, $32.60\%$, and $15.84\%$ average sum-rate gain over  PD-NOMA/OPT, PD-NOMA/SCA, and FRA/WF-PA, respectively.
	Note that,  FRA/WF-PA adopts an interference-free approach and for large interference thresholds at the CBS, all the transmitting nodes can use a high  transmit power. Due to these two factors, FRA/WF-PA can achieve high average sum-rate for large interference thresholds at the CBS. However,  the power allocations  in FRA/WF-PA are inversely proportional to the price of RRBs and the price of all the RRBs is increased for small  interference thresholds at the CBS. Hence, for the small  interference thresholds at CBS, the average sum-rate of FRA/WF-PA scheme is reduced severely, and RSMD achieves substantial sum-rate gain over FRA/WF-PA scheme.  We also observe that RSMD achieves improved average sum-rate than both PD-NOMA/OPT and PD-NOMA/SCA. Such an observation is expected from the discussion of Fig. 2b.
	%As explained in the discussion of Fig. 2b that in a given device-cluster,   RSMD  always achieves better E2E sum-rate than PD-NOMA/OPT. Furthermore,  our proposed  power allocation strategy adopts fractional programming based power control, which is superior than SCA based power control method.  For these two reasons,  RSMD  achieves average sum-rate gain over both  PD-NOMA/OPT and PD-NOMA/SCA. 
	Finally, we observe that the average sum-rate  gap between RSMD and PD-NOMA/OPT  is reduced for the large interference thresholds at the CBS. Such an observation can be explained by the following argument.  For the large interference thresholds at the CBS, all the nodes use (almost)  the maximum transmit power, and therefore, both RSMD and PD-NOMA/OPT can operate with the maximum achievable sum-rate of the system.  Intuitively, in such a case, the  average sum-rate gap between RSMD and PD-NOMA/OPT  is reduced.}
	%We conclude the RSMD has a clear merit over the benchmark schemes in the presence of interference constraints at the CBS.}
	
	%Fig. \ref{fig_5_a} depicts that RSMD has a cl
%	Note that, RSMD achieves sum-rate gain over PD-NOMA/OPT  by optimizing power allocations. For large interference thresholds at the CBS,  all the nodes use (almost)  the maximum transmit power  and in such a case,   the gain obtained by power allocations is small. Essentially, in such a case, the  average sum-rate gap between RSMD and PD-NOMA/OPT schemes is reduced.}
\begin{figure}
	\vspace*{-0.4cm}
	\begin{center}
		\subfloat[t][\textcolor{black}{E2E sum-rate vs. the number of PCVs in Algorithm \ref{Algorithm1}}]{
			\includegraphics[width=0.45\linewidth, draft=false]{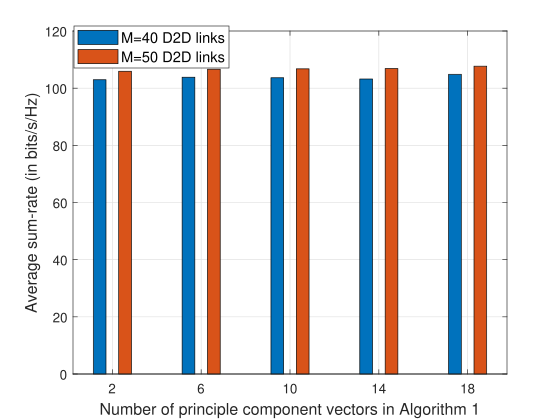}
			\label{fig_6_a}}
		\quad
		\subfloat[t][\textcolor{black}{E2E sum-rate vs. the number of D2D links}]{
			\includegraphics[width=0.45\linewidth, draft=false]{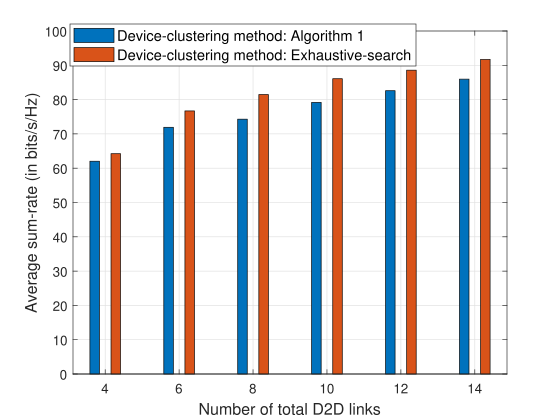}
			\label{fig_6_b}}
		
		\caption{\textcolor{black}{Performance of the proposed device-clustering algorithm for different PCVs and D2D links  considering $L=5$ eRRHs, $N=36$ RRBs, and $N_R=5$ device-clusters per eRRH} }
	\end{center}
	\vspace*{-0.2cm}
\end{figure}
 \textcolor{black}{Fig. \ref{fig_5_b} plots the average sum-transmit power of RSMD, PD-NOMA/OPT,  and FRA/WF-PA  while varying interference thresholds at CBS. Fig. 5b depicts that for the moderate to high  interference thresholds at CBS, FRA/WF-PA consumes high average sum-transmit power. The reason is two folds. First,  because of the reduced RRB per D2D link,  device transmitter of each D2D link in FRA/WF-PA  needs to increase transmit power to  improve  data rate. Moreover, such an increase of the transmit
power  is feasible when the interference threshold at CBS is large. These two factors allow the FRA/WF-PA scheme to improve the average sum-rate of the D2D links by increasing the average sum-transmit power  in the large interference threshold regime. However, as expected, the average sum-transmit power of FRA/WF-PA  is reduced for the small interference thresholds at the CBS.  Fig. \ref{fig_5_b}  depicts that for the small to moderate  interference thresholds, PD-NOMA/OPT consumes less  average sum-transmit power than RSMD, and for the small  interference thresholds, FRA/WF-PA consumes less average sum-transmit power than RSMD. Nonetheless, for both small and moderate interference thresholds,  RSMD achieves considerable average sum-rate gain over both PD-NOMA/OPT and FRA/WF-PA as confirmed from Fig. \ref{fig_5_a}. Therefore, despite having  small increase of  the average sum-transmit power, RSMD  has a clear merit  for both small and moderate  interference thresholds at the CBS.}

\vspace*{-0.3cm}
\subsection{Performance of the proposed device-clustering method}
\textcolor{black}{Fig. \ref{fig_6_a} plots the average sum-rate of RSMD for $40$ and $50$ D2D links while varying the number of PCVs in Algorithm \ref{Algorithm1}. Fig. \ref{fig_6_a} depicts that the average sum-rate of the proposed scheme is almost invariant for different number of  PCVs in Algorithm \ref{Algorithm1}.  Recall, an intrinsic property of the 2D-PCA is that the leading PCVs contain {most of the variability} of the original feature(s) \cite{PCA_Basic_1_Rev_1}. Usually, for the D2D links' channel matrices of the considered system, the higher-indexed PCVs have much small variances than the leading PCVs, and have little impact  on both the channel disparity  between two D2D links and the formation of device-clusters by using Algorithm \ref{Algorithm1}. 
	%	Using such a property we can readily conclude that 
	%	the most important information about the channel matrices for clustering the D2D links can be obtained from the first few PCVs. In fact, the higher number of PCVs do not significantly influence the channel disparity  between two D2D links, and hence, they have little impact on device-cluster formation. 
	As a result, \textit{on an average}, the sum-rate  is not  varied significantly as  the number of PCVs in Algorithm \ref{Algorithm1} is increased. Therefore,  our proposed device-clustering method can  work efficiently with the leading  PCVs. Such an observation  confirms the fact that  the  DUs of each D2D link need to upload only the first few PCVs to CBS to facilitate device-clustering, and thus, the required training overhead of our proposed Algorithm \ref{Algorithm1} is small.}

% Hence, the  required training overhead in the device-clustering phase of the proposed scheme can be substantially reduced.}

\textcolor{black}{Fig. \ref{fig_6_b} plots the average sum-rate of RSMD scheme for different number of D2D links. Here, the average sum-rate obtained by the proposed Algorithm \ref{Algorithm1} is compared with the average sum-rate obtained by an exhaustive-search  based device-clustering algorithm. An exhaustive-search  based device-clustering algorithm considers all the possible device-cluster combinations in the network, and as a result, it can achieve the optimal average sum-rate. However, the required computational complexity of the exhaustive-search is increased exponentially with the number of D2D links. In contrast, the complexity of  Algorithm \ref{Algorithm1} is increased at a quadratic rate with the number of D2D  links. Despite such a remarkable reduction of complexity, Algorithm \ref{Algorithm1} experiences a small average sum-rate loss compared to the exhaustive-search method. Fig. \ref{fig_6_b} shows that Algorithm \ref{Algorithm1} experiences maximum $8.8\%$ average sum-rate loss compared to the exhaustive-search method.  Accordingly, our proposed device-clustering method strikes a suitable balance between the performance and complexity, and it is efficient for a large-scale network.}

\vspace*{-0.2cm}
\section{Conclusion}
\vspace*{-0.2cm}
We  investigated the interference management problem of a D2D-integrated F-RAN architecture by applying an RS-based transmission strategy. A novel  resource allocation algorithm was developed by optimizing the device-clustering,  transmit power allocations, and assignment of the RRBs and eRRHs among the device-clusters.  Simulation results confirmed the following three observations: (i) \textcolor{black}{ the device-clustering phase of the proposed RSMD algorithm reduces the  required  training overhead and complexity  substantially at the cost of small average sum-rate loss}; (ii) our proposed RSMD algorithm achieves appreciable  average sum-rate gain over the contemporary PD-NOMA schemes; and (iii) our proposed RSMD algorithm also considerably outperforms  the non rate-splitting benchmark schemes. 

\appendices

\section{}

$\text{P0}$ will be an NP-hard optimization problem if the corresponding decision problem $\text{P0}$  is  NP-complete. The  decision problem of $\text{P0}$ can be expressed as follows. \textit{For a given set of power allocations, is it possible to find a set of device-clusters and the corresponding resource allocation as such the  $\text{C1}$-$\text{C5}$ constraints are satisfied?} To solve such a decision problem, we assume that certain D2D links in the network are pair-wise conflicting in a sense that they severely interfere with each other at any eRRH, and as a result, if they are clustered together, the effective rate of the resultant device-cluster is zero. Essentially, the overall decision problem can be divided into sub-problems. The first sub-problem is equivalent to the following  set-packing problem.  \textit{Can we cover the overall set of the D2D links  by $\lceil \frac{M}{2}\rceil$ disjoints sets as such no set  (i.e., device-cluster) contains two conflicting D2D links?} On the other hand,  given the device-clusters, the second sub-problem is described as follows.  \textit{ Can we obtain a feasible allocation of the available eRRHs and RRBs among these device-clusters as such the $\text{C2}$-$\text{C5}$ constraints are satisfied?} Clearly, the second sub-problem is equivalent to a multiple Knapsack problem. Since, both the set-packing and   multiple Knapsack problems are NP-complete, the overall decision problem of  $\text{P0}$ is also  NP-complete. As a result, $\text{P0}$ is an NP-hard optimization problem. \QED

\section{}

Algorithm \ref{Algorithm1} sequentially forms device-clusters $\mathcal{S}_1, \mathcal{S}_2, \cdots, \mathcal{S}_{|\mathcal{T}|}$. We first proof that $\mathcal{S}_j$ and $\mathcal{S}_{j+1}$ device-clusters are not Pareto-improvement pair, $\forall j$.  Without loss of generality, we assume that $\mathcal{S}_j=\{m,k\}$ and $\mathcal{S}_{j+1}=\{m',k'\}$, and both the $m$-th and $m'$-th D2D links prefer to be clustered with the $k$-th D2D link. Since the $m$-th D2D link is selected to choose its favorite cluster partner prior to the $m'$-th D2D link, as per the Step 6 of Algorithm \ref{Algorithm1}, the condition $\Delta_m> \Delta_{m'}$ is satisfied. As per the Step 5 of Algorithm \ref{Algorithm1}, $\Delta_m$ is defined as $\Delta_m=\min_{e \neq k}\mathtt{d}(\mathbb{B}_m, \mathbb{B}_e)-\mathtt{d}(\mathbb{B}_m, \mathbb{B}_k)$ and consequently, $\Delta_m \leq \mathtt{d}(\mathbb{B}_m, \mathbb{B}_{k'})-\mathtt{d}(\mathbb{B}_m, \mathbb{B}_k)$ is satisfied.  Meanwhile, for the $m'$-th D2D link,   we obtain $\Delta_{m'}=\mathtt{d}(\mathbb{B}_{m'}, \mathbb{B}_{k'})-\mathtt{d}(\mathbb{B}_{m'}, \mathbb{B}_k)$. Since $\Delta_m> \Delta_{m'}$ holds true, the inequality  $\mathtt{d}(\mathbb{B}_m, \mathbb{B}_{k'})-\mathtt{d}(\mathbb{B}_m, \mathbb{B}_k)>\mathtt{d}(\mathbb{B}_{m'}, \mathbb{B}_{k'})-\mathtt{d}(\mathbb{B}_{m'}, \mathbb{B}_k) \implies \mathtt{d}(\mathbb{B}_m, \mathbb{B}_{k'})+\mathtt{d}(\mathbb{B}_{m'}, \mathbb{B}_k) > \mathtt{d}(\mathbb{B}_m, \mathbb{B}_k)+\mathtt{d}(\mathbb{B}_{m'}, \mathbb{B}_{k'})$ holds true. As a result, as per the definition of Pareto-improvement pair, $\mathcal{S}_j$ and $\mathcal{S}_{j+1}$ device-clusters can not be Pareto-improvement pair, $\forall j$.  Moreover, if $\mathcal{S}_j$ and $\mathcal{S}_{j+k}$  device-clusters are Pareto-improvement pair, $\forall j$ and $ \exists k >1$, they must swap their members at the Step 11 of Algorithm \ref{Algorithm1}.  As a matter of fact,   the final device-cluster set of Algorithm \ref{Algorithm1} does not contain any Pareto-improvement pair. Since the output of Algorithm \ref{Algorithm1} does not contain any  Pareto-improvement pair, it is indeed Pareto-efficient. Therefore,  Algorithm \ref{Algorithm1} provides a Pareto-efficient solution to $\text{P1}$. \QED

\section{}

We first  provide the definition of a potential game \cite{Potential}.  Particularly, $\mathcal{G}$ will be  a potential game if it admits a potential function $W(\cdot,\cdot)$ such that $\forall \mathcal{S}_j  \in \mathcal{S}$ and $\forall \{ \mathbf{P}_{j,l_j,n}, \mathbf{Q}_{j,l_j,n}\}, \{\mathbf{P}_{j,l_j,n}', \mathbf{Q}_{j,l_j,n}'\} \in \Pi_j$, the following condition is satisfied.
\begin{equation}
\label{Potential_Def}
\begin{split}
&\Gamma_j\left(\{\mathbf{P}_{j,l_j,n}, \mathbf{Q}_{j,l_j,n}\} \bigg| \{\bm{x},\bm{y},\bm{\mu}\}\right)- \Gamma_j\left(\{\mathbf{P}_{j,l_j,n}', \mathbf{Q}_{j,l_j,n}'\} \bigg| \{\bm{x},\bm{y},\bm{\mu}\}\right)\\
&=W\left(\{\mathbf{P}_{j,l_j,n}, \mathbf{Q}_{j,l_j,n}\}, \{\mathbf{P}_{t,l_t,n'}, \mathbf{Q}_{t,l_t,n'}\}\right)- W\left(\{\mathbf{P}_{j,l_j,n}', \mathbf{Q}_{j,l_j,n}'\}, \{\mathbf{P}_{t,l_t,n'}, \mathbf{Q}_{t,l_t,n'}\}\right) \\
\end{split}
\end{equation}
where $\{\mathbf{P}_{t,l_t,n'}, \mathbf{Q}_{t,l_t,n'}\}$  denotes power allocations of the $\mathcal{S}_t$ device-cluster over the $l_t$-th eRRH and the $n'$-th RRB, and where $ \mathcal{S}_t \in \mathcal{S}\setminus \mathcal{S}_j$ and $n'\in \mathcal{N}_t$. Here, $\mathcal{N}_t$ is the set of allocated RRBs to the $\mathcal{S}_t$ device-cluster.

We now demonstrate that the considered NCPCG $\mathcal{G}$ admits a potential function given as  
%the following \textit{Proposition}
%\begin{proposition}
%	\label{Proposition_Potential}
%	The NCPCG $\mathcal{G}$ is an exact potential game with an exact potential function
\begin{equation}
\label{Potential_function}
W\left(\{\mathbf{P}_{j,l_j,n}, \mathbf{Q}_{j,l_j,n}\}, \{\mathbf{P}_{t,l_t,n'}, \mathbf{Q}_{t,l_t,n'}\}\right)=\sum_{j \in \mathcal{T}} \sum_{l \in\mathcal{L}} \sum_{n \in \mathcal{N}_{sc}} x_{j,l y_{j,l,n} \Xi_{j,l,n}}
\end{equation}
where  $\Xi_{j,l,n}$ is defined as
\begin{equation}
\label{Xi}
\begin{split}
\Xi_{j,l,n}=&R_{j,l_j,n}^{e2e}
-\bm{\sigma_j}^T\left[\sum_{i=1}^2  \mathbf{P}_{j,l,n}[i], \mathbf{P}_{j,l_j,n}[3], \sum_{i=1}^3  \mathbf{Q}_{j,l,n}[i] \right]^T\\
&-\mu_n\left(\sum_{i=1}^3 \mathbf{P}_{j,l,n}[i]+\sum_{i=1}^3 \mathbf{Q}_{j,l_j,n}[i]\right), \forall j,l,n.
\end{split}
\end{equation}
%\begin{table*}
%	\vspace*{-0.2cm}
%	\begin{normalsize} 

%	\end{normalsize}
%	\vspace*{-0.2cm}
%	\hrulefill
%\end{table*}

To justify that \eqref{Potential_function} is a potential function for NCPCG $\mathcal{G}$, recall that the assignments of the eRRHs and RRBs among the players (i.e., device-clusters) are given. Particularly, $x_{j,l}=1$ if $l=l_j$ and $x_{j,l}=0$, $\forall l \neq l_j$. Moreover, $y_{n,l,j}=1$ if $l=l_j$ and $n \in N_j$, and $y_{n,l,j}=0$, otherwise.
Similarly, for the $\mathcal{S}_t$ device-cluster, $x_{t,l}=1$ if $l=l_t$ and $x_{t,l}=0$, $\forall l \neq l_t$ where $t \in \mathcal{T}$ and $t \neq j$. Each device-cluster is allocated with orthogonal set of RRBs, and consequently, $\mathcal{N}_t \cap \mathcal{N}_j=\varnothing $ where $t \in \mathcal{T}$ and $t \neq j$.
Accordingly, we can rewrite the functions $	W\left(\{\mathbf{P}_{j,l_j,n}, \mathbf{Q}_{j,l_j,n}\}, \{\mathbf{P}_{t,l_t,n'}, \mathbf{Q}_{t,l_t,n'}\}\right)$ and $	W\left(\{\mathbf{P}_{j,l_j,n}', \mathbf{Q}_{j,l_j,n}'\}, \{\mathbf{P}_{t,l_t,n'}, \mathbf{Q}_{t,l_t,n'}\}\right)$ as
%\eqref{Potential_2} and \eqref{Potential_3}, respectively, at the top of current page.
\begin{equation}
\label{Potential_2}
\begin{split}
&	W\left(\{\mathbf{P}_{j,l_j,n}, \mathbf{Q}_{j,l_j,n}\}, \{\mathbf{P}_{t,l_t,n'}, \mathbf{Q}_{t,l_t,n'}\}\right)=\sum_{n \in \mathcal{N}_j} \Xi_{j,l_j,n}+ \sum_{\substack{t \in \mathcal{T} \\ t \neq j}}\sum_{\substack{n' \in \mathcal{N}_t\\ n' \neq n}} \Xi_{t,l_t,n'}\\
&=\Gamma_j\left(\{\mathbf{P}_{j,l_j,n}, \mathbf{Q}_{j,l_j,n}\} \bigg| \{\bm{x},\bm{y}, \bm{\mu}\}\right)+\sum_{t \in \mathcal{T}, t \neq j} \Gamma_t\left(\{\mathbf{P}_{t,l_t,n'}, \mathbf{Q}_{t,l_t,n'}\} \bigg| \{\bm{x},\bm{y},\bm{\mu}\}\right)
\end{split}
\end{equation}
%	\end{normalsize}
%	\vspace*{-0.2cm}
%	\hrulefill
%\end{table*}
%\begin{table*}
%	\vspace*{-0.2cm}
%	\begin{normalsize} 
and 
\begin{equation}
\label{Potential_3}
\begin{split}
&	W\left(\{\mathbf{P}_{j,l_j,n}', \mathbf{Q}_{j,l_j,n}'\}, \{\mathbf{P}_{t,l_t,n'}, \mathbf{Q}_{t,l_t,n'}\}\right)\\
&=\Gamma_j\left(\{\mathbf{P}_{j,l_j,n}', \mathbf{Q}_{j,l_j,n}'\} \bigg| \{\bm{x},\bm{y},\bm{\mu}\}\right)+\sum_{t \in \mathcal{T}, t \neq j} \Gamma_t\left(\{\mathbf{P}_{t,l_t,n'}, \mathbf{Q}_{t,l_t,n'}\} \bigg| \{\bm{x},\bm{y},\bm{\mu}\}\right).
\end{split}
\end{equation}
As a result, we obtain
\begin{equation}
\label{Potential_proof}
\begin{split}
& W\left(\{\mathbf{P}_{j,l_j,n}, \mathbf{Q}_{j,l_j,n}\}, \{\mathbf{P}_{t,l_t,n'}, \mathbf{Q}_{t,l_t,n'}\}\right)- W\left(\{\mathbf{P}_{j,l_j,n}', \mathbf{Q}_{j,l_j,n}'\}, \{\mathbf{P}_{t,l_t,n'}, \mathbf{Q}_{t,l_t,n'}\}\right)\\
&=\Gamma_j\left(\{\mathbf{P}_{j,l_j,n}, \mathbf{Q}_{j,l_j,n}\} \bigg| \{\bm{x},\bm{y},\bm{\mu}\}\right)- \Gamma_j\left(\{\mathbf{P}_{j,l_j,n}', \mathbf{Q}_{j,l_j,n}'\} \bigg| \{\bm{x},\bm{y},\bm{\mu}\}\right).
\end{split}
\end{equation}
Therefore, as per  the definition of the potential game, eq. \eqref{Potential_function} is a  potential function for the game $\mathcal{G}$, and consequently, $\mathcal{G}$ is a  potential game. Moreover, since every potential game posses at least one NE solution \cite{Potential}, $\mathcal{G}$ must posses at least one NE power allocation strategy. This completes the proof of Lemma 2. \QED

\section{}

The maximization problem in \eqref{BRS_I_obj} can be alternatively expressed as
\begin{equation}
\label{BRS_j_alterantive}
\begin{split}
&\max_{\mathbf{P} \geqq \bm{0}, \mathbf{Q} \geqq \bm{0}}\left(\sum_{n \in \mathcal{N}_j}(1-\lambda_{j,l_j,n})R_{j,l_j,n}^{(U)}+\lambda_{j,l_j,n}R_{j,l_j,n}^{(D)}\right.\\
&\left.\ -\sum_{n \in \mathcal{N}_j} \mu_n
\left(\sum_{i=1}^3  \mathbf{P}_{j,l_j,n}[i]+\sum_{i=1}^3\mathbf{Q}_{j,l_j,n}[i] \right)\right.\\
&\left.-\bm{\sigma_j}^T\left[\sum_{n \in\mathcal{N}_j}\sum_{i=1}^2  \mathbf{P}_{j,l_j,n}[i], \sum_{n \in\mathcal{N}_j} \mathbf{P}_{j,l_j,n}[3], \sum_{n \in\mathcal{N}_j}\sum_{i=1}^3  \mathbf{Q}_{j,l_j,n}[i] \right]^T\right).
\end{split}
\end{equation}	
where $\lambda_{j,l_j,n} \in (0,1)$ parameter. Owing to the flow conservation at the eRRH, the optimal value of $\lambda_{j,l_j,n}$ is obtained such that $\mathrm{R}_{j,l_j,n}^{(U)}=\mathrm{R}_{j,l_j,n}^{(D)}$ is satisfied,  $\forall j,n$. In fact, in such a case, eqs. \eqref{BRS_I_obj} and \eqref{BRS_j_alterantive} provide the same solution. Hence, in stead of solving  \eqref{BRS_I_obj} , we  iteratively solve \eqref{BRS_j_alterantive} for the given $\lambda_{j,l_j,n}$  and update $\lambda_{j,l_j,n}$ as such the data rates of both hops are balanced. Accordingly,  to prove \textit{Proposition 2}, it is sufficient to prove that for the given $\lambda_{j,l_j,n}$, eq. \eqref{BRS_j_alterantive} is equivalent to \eqref{BRS_II}. 

The required proof to show the equivalence between \eqref{BRS_II} and \eqref{BRS_j_alterantive}  has the similar steps of \cite[Theorem 3]{W_YU_II}.  By introducing two auxiliary variables $\{Z_{j,l_j,n}^{(i)}\}$ and $\{X_{j,l_j,n}^{(i)}\}$,  eq. \eqref{BRS_j_alterantive}  is equivalently  expressed as an optimization problem as \eqref{Auxilary_Optimization} at the top of the next page.
\setlength{\textfloatsep}{0pt}
\begin{table*}
	\begin{normalsize} 
		\begin{equation}
		\label{Auxilary_Optimization}
		\begin{split}
		\max_{\{\mathbb{P}, \mathbb{Q}, \mathbb{Z},\mathbb{X}\} \geqq 0} \mathcal{F}_j^{(0)} \triangleq
		&\sum_{n' \in \mathcal{N}_j}\left(\sum_{i=1}^2(1-\lambda_{j,l_j,n})\log_2\left(1+Z_{j,l_j,n}^{(i)}\right)+R_{j(1),l_j,n}^{(U,2)}+\sum_{i=1}^3\lambda_{j,l_j,n}\log_2\left(1+X_{j,l_j,n}^{(i)}\right)\right)\\
		&\qquad-\Psi_j(\mathbb{P},\mathbb{Q})\\
		& \text{s.t.} \begin{cases}
		& \text{C6:} \quad  \frac{A_{j,l_j,n}^{(i)}}{B_{j,l_j,n}^{(i)}} \geq Z_{j,l_j,n}^{(i)}, i=1,2 \\
		& \text{C6:} \quad  \frac{\widehat{A}_{j,l_j,n}^{(i)}}{\widehat{B}_{j,l_j,n}^{(i)}} \geq Z_{j,l_j,n}^{(i)}, i=1,2,3.
		\end{cases}
		\end{split}
		\end{equation}
	\end{normalsize}
	%\vspace*{-0.2cm}
	\hrulefill
\end{table*}
For the given $\{\mathbf{P,Q}\}$, eq. \eqref{Auxilary_Optimization} is a strict concave optimization problem of $\{\mathbf{Z,X}\}$, and hence, it has zero duality-gap. Assume that both $\{\mathbf{P,Q}\}$ are given, and $\left\{\{\Omega_i\}\right\}$ and $\left\{\varOmega_i \right\}$ are the Lagrangian multipliers correspond to the $\text{C6}$  and $\text{C7}$ constraints in \eqref{Auxilary_Optimization}, respectively. The Lagrangian function of \eqref{Auxilary_Optimization} is obtained as \eqref{Lagrangian} at the top of the next page.
\setlength{\textfloatsep}{0pt}
\begin{table*}
	\begin{normalsize} 
		\begin{equation}
		\label{Lagrangian}
		\begin{split}
		&\mathcal{L}_j=\sum_{n \in \mathcal{N}_j} \omega_{j,l_j,n}^{(1)} \left(\sum_{i=1}^2 \log\left(1+Z_{j,l_j,n}^{(i)}\right)+\log\left(1+\frac{ P_{j,l_j,n}^{(1,2)} \left|h_{j,l_j,n}^{(1)}\right|^2}{I_{c,l_j,n}}\right)\right)-\Psi_j(\mathbf{P,Q})\\
		& +\sum_{n \in \mathcal{N}_j} \omega_{j,l_j,n}^{(2)} \left(\sum_{i=1}^3 \log\left(1+X_{j,l_j,n}^{(i)}\right)\right)-\sum_{i=1}^2 \Omega_i\left(Z_{j,l_j,n}^{(i)}-\frac{A_{j,l_j,n}^{(i)}}{B_{j,l_j,n}^{(i)}}\right)-\sum_{i=1}^3 \varOmega_i\left(X_{j,l_j,n}^{(i)}-\frac{\widehat{A}_{j,l_j,n}^{(i)}}{\widehat{B}_{j,l_j,n}^{(i)}}\right).
		\end{split}
		\end{equation}
	\end{normalsize}
	\hrulefill
\end{table*}
The Lagrangian dual-function is defined as $\mathcal{L}_j^*= \max_{\mathbf{Z,X}} \mathcal{L}_j|_{\Omega_i^{}=\Omega_i^{*}, \varOmega_i=\varOmega_i^{*}}$ where $\{\Omega_i^{*}\}$ and $\{\varOmega_i^*\}$ are the optimal Lagrangian multipliers. We can justify that at the optimality of \eqref{Lagrangian}, $\Omega_i^{*}=\frac{\omega_{j,l,n}^{(1)}}{1+Z_{j,l_j,n}^{(i)^*}}, \forall i=1,2$, and $\varOmega_i^{*}=\frac{\omega_{j,l,n}^{(2)}}{1+X_{j,l_j,n}^{(i)^*}}, \forall i=1,2,3$,  are satisfied. Here, $\{Z_{j,l_j,n}^{(i)^*}\}$ and $\{X_{j,l_j,n}^{(i)^*}\}$ are the optimal values of $\{Z_{j,l_j,n}^{(i)}\}$ and $\{X_{j,l_j,n}^{(i)}\}$, respectively. As depicted from \eqref{Lagrangian}, $Z_{j,l_j,n}^{(i)^*}=\frac{A_{j,l_j,n}^{(i)}}{B_{j,l_j,n}^{(i)}}$ and $X_{j,l_j,n}^{(i)^*}=\frac{\widehat{A}_{j,l_j,n}^{(i)}}{\widehat{B}_{j,l_j,n}^{(i)}}$ must hold true; otherwise, $\mathcal{L}^*$ would approach infinity.  Hence,  after some straightforward manipulations, $\mathcal{L}_j^*$ is obtained as
\begin{equation}
\label{Lagrangian_dual_function}
\mathcal{L}_j^*=\mathcal{F}_j^{(1)}\left(\mathbf{Z}, \mathbf{X}\right) + \mathcal{F}_j^{(2)}\left(\mathbf{P}, \mathbf{Q},\mathbf{Z}, \mathbf{X}\right)-\Psi_j\left(\mathbf{P},\mathbf{Q}\right)
\end{equation}
where $\mathcal{F}_j^{(1)}\left(\mathbf{Z}, \mathbf{X}\right)$, $\mathcal{F}_j^{(2)}\left(\mathbf{P}, \mathbf{Q},\mathbf{Z}, \mathbf{X}\right)$, and $\Psi_j\left(\mathbf{P},\mathbf{Q}\right)$ are defined in \eqref{F_j_1}, \eqref{F_j_2}, and \eqref{Psi_j}, respectively.
Using a primal-decomposition, we obtain $\max_{\{\mathbf{P}, \mathbf{Q}, \mathbf{Z},\mathbf{X}\} \in \mathcal{C}} \mathcal{F}_j^{(0)}=\max_{\{\mathbf{P}, \mathbf{Q}\} \geqq 0}\left(\max_{\{ \mathbf{Z},\mathbf{X}\} \in \mathcal{C}} \mathcal{F}_j^{(0)}\right)$ where $\mathcal{C}=\left\{\mathbf{P,Q,Z,X}\bigg| \text{C6}\right\}$.  Owing to the strict concavity of \eqref{Auxilary_Optimization} for the given $\{\mathbf{P,Q}\}$, $\max_{\{ \mathbf{Z},\mathbf{X}\} \in \mathcal{C}} \mathcal{F}_j^{(0)}=\max_{\{ \mathbf{Z},\mathbf{X}\} \geqq 0} \mathcal{L}_j^* $ is satisfied. As a result, we obtain $\max_{\{\mathbf{P}, \mathbf{Q}, \mathbf{Z},\mathbf{X}\} \in \mathcal{C}} \mathcal{F}_j^{(0)}=\max_{\{\mathbf{P}, \mathbf{Q}, \mathbf{Z}, \mathbf{X}\} \geqq 0}  \mathcal{L}_j^*$.  Note that \eqref{BRS_II} and \eqref{Lagrangian_dual_function} are same.
Consequently, the optimization problem given by \eqref{BRS_j_alterantive} can be solved equivalently  by maximizing  \eqref{BRS_II} with respect to $\{\mathbf{P,Q,Z,X}\}$.  This completes the proof of \textit{Proposition 2}. \QED

\section{}

Based on \cite[Proposition 1]{W_YU_II}, it can be readily verified that by  solving \eqref{Outer} and \eqref{Quadratic_Inner} alternately, a  non-decreasing sequence of $\{\Gamma_j\}$ is obtained, $\forall j \in \mathcal{T}$. Hence, the solution obtained by  solving  \eqref{Outer} and \eqref{Quadratic_Inner} alternately can monotonically improve the device-cluster's payoff.  Recall, because of the maximum transmit power limit of the CH DU and eRRH, each device-cluster's payoff is bounded above.  Moreover,  $\mathcal{G}$ is a potential game, and every potential game exhibits the finite improvement property \cite[Lemma 2.3]{Potential}. Particularly, an algorithm that monotonically improves the players' payoffs, must converge to an NE strategy. Therefore, the solution obtained by alternatively solving  \eqref{Outer} and \eqref{Quadratic_Inner}  converges to an NE  power allocation strategy of the device-clusters. \QED

\section{}

Every NE strategy of a potential game is a local maximizer of the potential function associated with the game \cite{Potential}. As per \textit{Lemma} \ref{Corollary_Potenital_2}, Algorithm \ref{Algorithm2} converges to an NE strategy for the  potential game, $\mathcal{G}$, and hence, it obtains a local-optimal solution to the  potential function $W(\cdot, \cdot)$ given in \eqref{Potential_function}. We  observe that in \eqref{Potential_function}, the parameters $\{\sigma_{j,1},\sigma_{j,2}, \sigma_{j,3}\}$ are in fact the Lagrangian multipliers correspond to the constraint $\text{C4}$ of $\text{P2.1}$\footnote{For this reason, we utilize a sub-gradient method to update $\{\sigma_{j,1},\sigma_{j,2}, \sigma_{j,3}\}$ in Algorithm \ref{Algorithm2}.}, and consequently, $ W(\cdot, \cdot)$ is also the partial Lagrangian dual-function of $\text{P2.1}$. Therefore, a local maximizer of $W(\cdot, \cdot)$ must satisfy the first-order optimality condition for  $\text{P2.1}$. As a result, Algorithm \ref{Algorithm2} converges at least to a local-optimal solution  to $\text{P2.1}$. \QED

\section{}

Based on \cite[Theorem 1]{Bilinear}, the convergence of Algorithm  \ref{Algorithm3} to a local-optimal solution to \eqref{Optimization_Leader_1} can be established. For the completeness of the proof, we denote the value of the objective function of \eqref{Optimization_Leader_1} as $\mathtt{V}_L(\mathbf{x}, \mathbf{y})$. Moreover, we denote $\left\{\mathbf{x}^{(t+1)}, \mathbf{y}^{(t+1)}\right\}$ and  $\left\{\mathbf{x}^{(t)}, \mathbf{y}^{(t)}\right\}$ as the output of Algorithm  \ref{Algorithm3} at the $(t+1)$-th and $t$-th iterations, respectively. Since, in Step 5 of Algorithm \ref{Algorithm3}, eq. \eqref{Optimization_Leader_2} is optimally  solved by using the Hungarian algorithm \cite{Hungarian},  $\mathtt{V}_L \left(\mathbf{x}^{(t)}, \mathbf{y}^{(t)}\right) \leq \mathtt{V}_L \left(\mathbf{x}^{(t+1)}, \mathbf{y}^{(t)}\right)$ is satisfied. Moreover, Step 7 of Algorithm \ref{Algorithm3} optimally solves \eqref{Optimization_Leader_3}, and consequently,  $\mathtt{V}_L \left(\mathbf{x}^{(t+1)}, \mathbf{y}^{(t)}\right) \leq \mathtt{V}_L \left(\mathbf{x}^{(t+1)}, \mathbf{y}^{(t+1)}\right)$ is satisfied. Therefore, we obtain $\mathtt{V}_L \left(\mathbf{x}^{(t)}, \mathbf{y}^{(t)}\right) \leq \mathtt{V}_L \left(\mathbf{x}^{(t+1)}, \mathbf{y}^{(t+1)}\right)$. In other words, Algorithm  \ref{Algorithm3} generates a sequence of $\left(\mathbf{x}, \mathbf{y}\right)$ that
non-decreasingly improves the objective function of \eqref{Optimization_Leader_1}. Since the optimal solution to \eqref{Optimization_Leader_1} is bounded above, Algorithm  \ref{Algorithm3}
must converge to a local-optimal solution to \eqref{Optimization_Leader_1}. \QED

\section{}

We first justify that the proposed RSMD algorithm achieves the converged resource allocation. As described in Section IV. C, the price of the $n$-th RRB, $\mu_n$, is searched  between $\{\mu_n^{(low)}\}$ and $\{\mu_n^{(high)}\}$  by using a bi-section search method. On the other hand, the transmit power allocations obtained by \eqref{CH_DU_power_1st_stream}-\eqref{CH_DU_power_3rd_stream} and \eqref{relay_CU_power_1st_stream}-\eqref{relay_CU_power_3rd_stream} are both inversely proportional to the prices of the RRBs. Such an observation leads to the fact that when $\{\mu_n^{(low)}\}$ and $\{\mu_n^{(high)}\}$ are sufficiently small and large, respectively,  the bi-section search method of updating the RRBs' prices will always converge to certain $\{\mu_n^*\}$ such that $|I_n^{(up)}-I_{th}|$ approaches a small value, $\forall n \in \mathcal{N}_{sc}$.  Moreover, as the prices of the RRBs are converged, the power allocations obtained from  Step 12 of the RSMD algorithm must be stable as well. Therefore, RSMD algorithm achieves the converged resource allocation. Let $\{\bm{P}^*, \bm{Q}^*\}$ be the converged power allocations, and $\bm{\mu}^{*}$ be converged prices of the RRBs. Next, we justify that such a converged resource allocation is the SE for the considered Stackelberg game.

Recall, the main goal of the CBS in the considered Stackelberg game is to satisfy the uplink  interference constraints. Therefore, as long as the uplink interference constraints are satisfied, CBS  would not alter  prices of the RRBs from $\bm{\mu}^{*}$ to other values.  On the other hand, as per \textit{Lemma } \ref{Corollary_Potenital_2}, for the given RRBs' prices and the resource assignments, the converged power allocations obtained by Algorithm \ref{Algorithm2} provide the NE outcome for the device-clusters. Moreover, as per \textit{Proposition} \ref{Proposition_Potential_Optimality}, such power allocations also ensure social-optimality for the device-clusters. Therefore,  as long as the RRBs' prices and the resource assignments do not change, device-clusters can not  unilaterally  improve their utility (or payoff) by deviating from the converged power allocations obtained by Algorithm \ref{Algorithm2}. Note that, under the converged RRBs' prices, $\bm{\mu}^{*}$,  $\{\bm{P}^*, \bm{Q}^*\}$ are the converged power allocations of the device-clusters. Therefore, given $\bm{\mu}^{*}$, the device-clusters  will not have  any incentive to change the power allocations from $\{\bm{P}^*, \bm{Q}^*\}$ to other values, and given the device-clusters' power allocations are not changed, the CBS has no incentive to change the prices of the RRBs from $\bm{\mu}^{*}$ to other values. Therefore, as per the definition of SE, RSMD algorithm converges to the SE resource allocation strategy. \QED

%\vspace*{-0.2cm}

\end{spacing}

\end{document}